\documentclass[twocolumn,showpacs,preprintnumbers,amsmath,amssymb]{revtex4}

\usepackage{graphicx}
\usepackage{dcolumn}
\usepackage{bm}
\usepackage{multirow}
\usepackage{color}

\newcommand{\br}{\mbox{\boldmath$r$}}
\newcommand{\bj}{\mbox{\boldmath$j$}}
\newcommand{\svec}[1]{{\mbox{\boldmath$#1$}}}
\newcommand{\balp}{\mbox{\boldmath$\alpha$}}
\newcommand{\bp}{{\mathbf{p}}}
\newcommand{\bnab}{\mbox{\boldmath$\nabla$}}
\newcommand{\bmu}{\mbox{\boldmath$\mu$}}
\newcommand{\bSig}{\mbox{\boldmath$\Sigma$}}

\begin{document}

\title{Relativistic description of magnetic moments in nuclei with doubly closed shells plus or minus one nucleon}

\author{J. Li}
 \affiliation{College of Physics, Jilin University, Changchun 130012, China}
\author{J. X. Wei}
  \affiliation{School of Physics and Nuclear Energy Engineering, Beihang University,Beijing 100191, China}
\author{J. N. Hu}
  \affiliation{State Key Laboratory of Nuclear Physics and Technology, School of Physics, Peking University, Beijing 100871, China}

\author{P. Ring}
 \affiliation{State Key Laboratory of Nuclear Physics and Technology, School of Physics, Peking University, Beijing 100871, China}
 \affiliation{Physik Department, Technische Universit\"{a}t M\"{u}nchen, D-85747 Garching, Germany}
\author{J. Meng}
 \email{mengj@pku.edu.cn}
 \affiliation{State Key Laboratory of Nuclear Physics and Technology, School of Physics, Peking University, Beijing 100871, China}
 \affiliation{School of Physics and Nuclear Energy Engineering, Beihang University, Beijing 100191, China}
  \affiliation{Department of Physics, University of Stellenbosch, Stellenbosch, South Africa}

\date{\today}

\begin{abstract}
Using the relativistic point-coupling model with density functional PC-PK1, the magnetic moments of the nuclei $^{207}$Pb, $^{209}$Pb, $^{207}$Tl and $^{209}$Bi with a $jj$ closed-shell core $^{208}$Pb are studied on the basis of relativistic mean field (RMF) theory. The corresponding time-odd fields, the one-pion exchange currents, and the first- and second-order corrections are taken into account. The present relativistic results reproduce the data well. The relative deviation between theory and experiment for these four nuclei is $6.1\%$ for the relativistic calculations and somewhat smaller than the value of $13.2\%$ found in earlier non-relativistic investigations. It turns out that the $\pi$ meson is important for the description of magnetic moments, first by means of one-pion exchange currents and second by the residual interaction provided by the $\pi$ exchange.
\end{abstract}

\maketitle

\section{Introduction}

The nuclear magnetic moment is an important observable in nuclear physics. It provides rich information about nuclear structure, serves as a stringent test of nuclear models, and has
attracted the attention of nuclear physicists since the early days~\cite{Blin-Stoyle1956,Arima1984,Castel1990,Talmi2005a}. The theoretical description of nuclear magnetic moments has a long-standing
problem. In the last decades, many successful nuclear structure
models have been developed. However, the application of these
models for nuclear magnetic moments is still not satisfactory.

In the extreme single-particle shell model, the magnetic moment of
an odd-$A$ nuclei is carried only by one valence nucleon (valence-nucleon approximation), which
leads to the well-known Schmidt values. It was observed in the
early 1950s~\cite{Blin-Stoyle1953}, however, that almost all nuclear magnetic
moments are sandwiched between the two Schmidt lines~\cite{Schmidt1937}. Therefore, considerable efforts have
been made to explain the deviations of the nuclear magnetic moments
from the Schmidt values, which can be contributed from meson exchange current (MEC, i.e.,
the exchange of charged meson) and configuration mixing (CM, or core polarization, i.e.,
the correlation not included in the mean-field approximation)~\cite{Towner1987,Arima1987,Arima2011}.

In 1954, Arima and Horie~\cite{Arima1954} pointed out a very distinct
difference between the following two groups of nuclei: The cores
of the nuclei in the first group ($^{16}$O and $^{40}$Ca cores) are $LS$-closed,
i.e., the spin-orbit partners $j = l \pm \frac{1}{2}$
of the core are completely occupied. Therefore they are expected not to be
excited strongly by an external field with $M1$ character. On the other hand,
the cores of the nuclei in the second group (like $^{208}$Pb) are $jj$-closed,
i.e., one of the spin-orbit partners is open, and an $M1$
external field can strongly excite core nucleons to the empty
spin-orbit partner. This $M1$ giant resonance state of the core
can be momentary excited by the interaction with the valence
nucleon. This is the idea of first-order configuration mixing,
which is also called the Arima-Horie effect~\cite{Talmi2005}. It explains not only the difference between these two groups of nuclei, but also deviations of magnetic moments from the Schmidt lines for many nuclei~\cite{Arima1954,Arima1954b,Hiroshi1958}.

In the late 1960s, it became clear after many calculations that the first-order effect is not enough to explain the large deviations from the Schmidt values in high spin states, especially $^{209}$Pb.
Pion exchange is very important to understand nuclear magnetic moments, as was first
pointed out by Miyazawa in 1951~\cite{Miyazawa1951} and by Villars in 1952~\cite{Villars1952}. This correction changes the gyromagnetic ration of orbital angular momentum of a nucleon in the nucleus~\cite{Chemtob1969} and improves the agreement between theoretical and observed values~\cite{Hyuga1980}. Furthermore, the second-order configuration mixing had been taken into account firstly in Refs.~\cite{Ichimura1965,Mavromatis1966,Mavromatis1967}, which are also important to understand the deviations as well as meson exchange current~\cite{Shimizu1974,Towner1983}.

In the past decades, relativistic mean field (RMF) theory has
been successfully applied to the analysis of nuclear structure over
the whole periodic table, from light to superheavy nuclei with a few
universal parameters~\cite{Reinhard1989,Ring1996,Vretenar2005,Meng2006}. However, relativistic descriptions of nuclear magnetic moments are mostly restricted to $LS$ closed-shell nuclei $\pm1$ nucleon. It is known for some time that in straightforward applications of the relativistic single-particle model, where only sigma and the time-like component of the vector mesons were considered, the predicted isoscalar magnetic moments were significantly larger than the observed
values~\cite{Miller1975,Serot1981}. This is
because the reduced Dirac effective nucleon mass ($M^*\sim 0.6M$) enhances the relativistic effect on the electromagnetic current~\cite{McNeil1986}. It was pointed out that the valence-nucleon approximation is wrong in the relativistic calculation~\cite{Serot1992}.

After the introduction of vertex corrections by the definition of
effective single-particle currents in nuclei, i.e., the
``back-flow" effect in the framework of a relativistic
extension of Landau's Fermi-liquid theory~\cite{McNeil1986} or
by a random phase approximation (RPA) type summation of p-h and
p-$\bar{n}$ bubbles in relativistic Hartree
approximation~\cite{Ichii1987a,Shepard1988,Furnstahl1988}, or by the
consideration of non-zero space-like components of vector
mesons in a self-consistent deformed RMF
theory~\cite{Hofmann1988,Furnstahl1989,Yao2006,Li2009,Li2009b}, the isoscalar
magnetic moments of odd nuclei in the direct vicinity of $LS$
closed shells could be reproduced rather well.

Unfortunately, these effects cannot remove the discrepancy existing
in isovector magnetic moments. To eliminate
this discrepancy, one-pion exchange current corrections have
been included in the relativistic model~\cite{Morse1990,Li2011b},
which was found to be significant. However they lead to
a larger disagreement with data. Recently, the second-order configuration mixing has been included in the fully self-consistent relativistic theory, and greatly improves the description of isovector magnetic moment~\cite{Li2011,Wei2012}.

In Ref.~~\cite{Furnstahl1987}  magnetic moments of $jj$ closed-shell $\pm1$ nuclei near $^{208}$Pb have been studied in RMF theory including the contribution from the core. The corresponding results show an improvement in comparison with the valence-nucleon approximation. On the other hand, meson exchange currents and configuration mixing, especially the first-order configuration mixing, are very important for the description of the magnetic moment of such nuclei.

In view of these facts we present in this manuscript an investigation of magnetic moments of the $jj$ closed-shell $\pm1$ nuclei, $^{207}$Pb, $^{209}$Pb, $^{207}$Tl, and $^{209}$Bi in a relativistic framework, based on the magnetic moments derived from RMF theory with time-odd fields, including one-pion exchange currents, and first- and second-order configuration mixing corrections. For these calculations, the relativistic point-coupling (PC) model will be adopted. In Sec. II we outline briefly the theoretical framework of this model and we present the definition of the  magnetic moment operator, the one-pion exchange currents and configuration mixing diagramms in first and second order. The numerical details are given in Sec. III. The calculations are described and the results are discussed in Sec. IV. Finally, Sec. V contains a brief summary and a perspective.

\section{The relativistic framework}
\subsection{Relativistic mean field theory}
The basic building blocks of RMF theory with point-couplings are the vertices
\begin{equation}\label{Block}
  (\bar \psi{\cal O}\Gamma\psi),\quad {\cal O}\in\{1,\vec\tau \},\quad
  \Gamma\in\{1,\gamma_\mu,\gamma_5,\gamma_5\gamma_\mu,\sigma_{\mu\nu}\},
\end{equation}
where $\psi$ is the Dirac spinor field, $\vec{\tau}$ is the isospin Pauli matrix, and $\Gamma$ generally denotes the $4\times4$ Dirac matrices. There are 10 such building blocks characterized by their transformation properties in isospin and in Minkowski space. We adopt arrows to indicate vectors in isospin space and bold type for the
space vectors. Greek indices $\mu$ and $\nu$ run over the Minkowski indices 0, 1, 2, and 3.

A general effective Lagrangian can be written as a power
series in $\bar \psi{\cal O}\Gamma\psi$ and their derivatives. In the present work, we
start with the following Lagrangian density~\cite{Buvenich2002}:
\begin{eqnarray}\label{EQ:LAG}
 {\cal L} &=& \bar\psi(i\gamma_\mu\partial^\mu-m)\psi-\frac{1}{4}F^{\mu\nu}F_{\mu\nu}-e\frac{1-\tau_3}{2}\bar\psi\gamma^\mu\psi
A_\mu\nonumber\\
    & & -\frac{1}{2}\alpha_S(\bar\psi\psi)(\bar\psi\psi)
        -\frac{1}{2}\alpha_V(\bar\psi\gamma_\mu\psi)(\bar\psi\gamma^\mu\psi)\nonumber\\
    & & -\frac{1}{2}\alpha_{TV}(\bar\psi\vec{\tau}\gamma_\mu\psi)(\bar\psi\vec{\tau}
        \gamma^\mu\psi)\nonumber\\
    & & -\frac{1}{3}\beta_S(\bar\psi\psi)^3-\frac{1}{4}\gamma_S(\bar\psi\psi)^4-\frac{1}{4}
    \gamma_V[(\bar\psi\gamma_\mu\psi)(\bar\psi\gamma^\mu\psi)]^2\nonumber\\
    & & -\frac{1}{2}\delta_S\partial_\nu(\bar\psi\psi)\partial^\nu(\bar\psi\psi)
        -\frac{1}{2}\delta_V\partial_\nu(\bar\psi\gamma_\mu\psi)\partial^\nu(\bar\psi\gamma^\mu\psi) \nonumber \\
    & & -\frac{1}{2}\delta_{TV}\partial_\nu(\bar\psi\vec\tau\gamma_\mu\psi)
       \partial^\nu(\bar\psi\vec\tau\gamma_\mu\psi).
\end{eqnarray}
There are totally 9 coupling constants, $\alpha_S$, $\alpha_V$,
$\alpha_{TV}$, $\beta_S$, $\gamma_S$, $\gamma_V$, $\delta_S$,
$\delta_V$, and $\delta_{TV}$. The subscripts $S$, $V$, and $T$
respectively indicate the symmetries of the couplings, i.e., $S$
stands for scalar, $V$ for vector, and $T$ for isovector.

Using the mean-field approximation and the ¡°no-sea¡±
approximation, one finds the energy density functional for a nuclear system,
\begin{equation}\label{Eq:Energy-PC}
E_{\rm DF}[\hat{\rho}]
= \int d^3\bm{r}~\mathcal{E}(\bm{r}),
\end{equation}
with the energy density
\begin{equation}\label{EDF}
 \mathcal{E}
  = \mathcal{E}_{\rm kin}(\bm{r})
    +  \mathcal{E}_{\rm int}(\bm{r})
    +  \mathcal{E}_{\rm em}(\bm{r}),
\end{equation}
which is composed of a kinetic part
\begin{equation}
   \mathcal{E}_{\rm kin}(\bm{r})
   =\sum\limits_{k=1}^A\psi^\dagger_k(\bm{r})(\balp\cdot\bp+\beta m_N-m_N)
   \psi_k(\bm{r}),
\end{equation}
where the sum over $k$ runs over the occupied orbits in the Fermi see (no-sea approximation), and an interaction part
\begin{eqnarray}
 \label{E12}
 \mathcal{E}_{\rm int}(\bm{r})
 &=& \frac{\alpha_S}{2}\rho_S^2+\frac{\beta_S}{3}\rho_S^3
    + \frac{\gamma_S}{4}\rho_S^4+\frac{\delta_S}{2}\rho_S\triangle \rho_S \nonumber\\
 &&
    + \frac{\alpha_V}{2}j_\mu j^\mu + \frac{\gamma_V}{4}(j_\mu j^\mu)^2 +
       \frac{\delta_V}{2}j_\mu\triangle j^\mu  \nonumber\\
 && +  \frac{\alpha_{TV}}{2}\vec j^{\mu}_{TV}\cdot(\vec j_{TV})_\mu+\frac{\delta_{TV}}{2}
    \vec j^\mu_{TV}\cdot\triangle(\vec j_{TV})_{\mu}, \nonumber\\
\end{eqnarray}
with the local densities and currents
  \begin{subequations}\label{Eq:dencur-PC}
  \begin{eqnarray}
  \rho_S(\bm{r})&=&\sum_{k=1}^A\bar\psi_k(\bm{r})\psi_k(\bm{r}),\\
  j^\mu(\bm{r})&=&\sum_{k=1}^A\bar\psi_k(\bm{r})\gamma^\mu\psi_k(\bm{r}),\\%
  \vec j^{\mu}_{TV}(\bm{r})
              &=&\sum_{k=1}^A\bar\psi_k(\bm{r})\gamma^\mu\vec\tau
               \psi_k(\bm{r}),
 \end{eqnarray}
 \end{subequations}
and an electromagnetic part
\begin{equation}
   \mathcal{E}_{\rm em}(\bm{r})
   = \frac{1}{4}F_{\mu\nu}F^{\mu\nu}-F^{0\mu}\partial_0A_\mu+e A_\mu j^\mu_p.
\end{equation}
Minimizing the energy density functional Eq.~(\ref{Eq:Energy-PC}) with respect to $\bar{\psi}_k$, one obtains the Dirac equation for the single nucleons,
\begin{equation}\label{Eq:Dirac-PC}
  [-i\balp\cdot\bnab+\beta\gamma_\mu V^\mu+\beta(m+S)]\psi_k(\bm{r})=\varepsilon_k\psi_k(\bm{r}).
\end{equation}
The single-particle effective Hamiltonian contains local scalar
$S(\bm{r})$ and vector $V^\mu(\bm{r})$ potentials given by
\begin{equation}\label{Eq:potential-PC}
S(\bm{r})    =\Sigma_S, \quad
V^\mu(\bm{r})=\Sigma^\mu+\vec\tau\cdot\vec\Sigma^\mu_{TV},
\end{equation}
where the self-energies are given in terms of various densities,
  \begin{subequations}
  \begin{eqnarray}
  \Sigma_S           &=&\alpha_S\rho_S+\beta_S\rho^2_S+\gamma_S\rho^3_S+\delta_S\triangle\rho_S,\\
  \Sigma^\mu         &=&\alpha_Vj^\mu_V +\gamma_V (j^\mu_V)^3
                       +\delta_V\triangle j^\mu_V + e A^\mu,\\
  \vec\Sigma^\mu_{TV}&=& \alpha_{TV}\vec j^\mu_{TV}+\delta_{TV}\triangle\vec j^\mu_{TV}.
 \end{eqnarray}
 \end{subequations}
For the ground state of an even-even nucleus one has
time-reversal symmetry and the spacelike parts of the
currents ${\bm{j(r)}}$ in Eq.~(\ref{Eq:dencur-PC}) as well as the vector
potential $\bm{V(r)}$ in Eq.~(\ref{Eq:potential-PC}), i.e., the time-odd fields, vanish. However, in
odd-$A$ nuclei, the odd nucleon breaks the time-reversal
invariance, and time-odd fields give rise to a nuclear magnetic potential, which is very
important for the description of magnetic moments~\cite{Hofmann1988,Furnstahl1989}.
Moreover, because of charge conservation in nuclei, only the third component of
isovector potentials, $\vec{\Sigma}^\mu_{TV}$, contributes. The Coulomb field $A_0({\bm r})$
is determined by Poisson's equation and we neglect in the applications
the magnetic part ${\bm{A(r)}}$ of the electro-magnetic potential.

The relativistic residual interaction is given by the second derivative of the energy density
functional $E(\hat{\rho})$ with respect to the density matrix
\begin{equation}\label{eq:res.int.}
    V_{\alpha\beta\alpha'\beta'}
    = \frac{\delta^2E(\hat{\rho})}{\delta\hat{\rho}_{\alpha\beta}\delta\hat{\rho}_{\alpha'\beta'}}.
\end{equation}
More details can be found in Refs.~\cite{Niksic2005,Daoutidis2009}.

Although, because of parity conservation, the pion meson does not contribute to the ground state in RMF theory, it playa an important role in spin-isospin excitations and is usually included in relativistic RPA and QRPA calculations of these modes~\cite{DeConti1998,Paar2004}. The widely used pion-nucleon vertex reads in its pseudovector coupling form,
\begin{equation}\label{Eq:lag-PN}
     \mathcal {L}_{\pi N}
    = -\frac{f_\pi}{m_\pi}\bar{\psi}\gamma^\mu\gamma_5\vec{\tau}\psi\cdot\partial_\mu\vec{\pi},
\end{equation}
where $\vec{\pi}({\bm r})$ is the pion field, $f_\pi$ is the $\pi$-nucleon coupling constant and $m_\pi$ the pion mass.

\subsection{The magnetic moment operator}

The effective electromagnetic current operator used to describe the
nuclear magnetic moment is given by~\cite{Furnstahl1989,Yao2006,Li2009}
\begin{equation}\label{electromagnetic-current}
 \hat{J}^\mu(x) =
                   Q\bar{\psi}(x)\gamma^\mu\psi(x)+\frac{\kappa}{2M}\partial_\nu
                   [\bar{\psi}(x)\sigma^{\mu\nu}\psi(x)],
\end{equation}
where $Q\equiv\dfrac{e}{2}(1-\tau_3)$ is the nucleon charge,
$\sigma^{\mu\nu}=\dfrac{i} {2} [\gamma^\mu,\gamma^\nu]$ is the antisymmetric tensor operator, and
$\kappa$ the free anomalous gyromagnetic ratio of the nucleon,
$\kappa_p=1.793$ and $\kappa_n=-1.913$. In Eq.~(\ref{electromagnetic-current}), the first term gives the Dirac current and second term is the so-called anomalous current. The nuclear dipole magnetic moment in units of the nuclear magneton $\mu_N=\dfrac{e\hbar}{2Mc}$,
is determined by
\begin{subequations}
\begin{eqnarray}\label{magnetic-moment}
  \mbox{\boldmath{$\mu$}}
            &=& \frac{1}{2\mu_N}\int d^3r
                \br\times\langle g.s. \vert \hat\bj(\br)\vert g.s. \rangle,\\
            &=& \int d\mathbf{r}\,[\frac{Mc^2}{\hbar c}Q\psi^+(\mathbf{r})
             \mathbf{r}\times\balp\psi(\mathbf{r})
                 +\kappa\psi^+(\mathbf{r})\beta\bSig\psi(\mathbf{r})],\nonumber\\
\end{eqnarray}
\end{subequations}
where $\hat\bj(\br)$ is the operator of space-like components of the effective electromagnetic current.
The first term in above equation gives the Dirac magnetic moment, and the second term gives the anomalous magnetic moment.

Therefore, in the relativistic theory, the nuclear magnetic moment operator, in units of the nuclear magneton, is
given by
\begin{eqnarray}
 \label{mm-operator}
  \mbox{\boldmath{$\hat{\mu}$}}
             &=& \frac{Mc^2}
             {\hbar c}Q\bm {r}\times\bm{\alpha}
             +\kappa\beta\bm{\Sigma}.
\end{eqnarray}

\subsection{The one-pion exchange current}

\begin{figure}
\centerline{
\includegraphics[width=7.5cm]{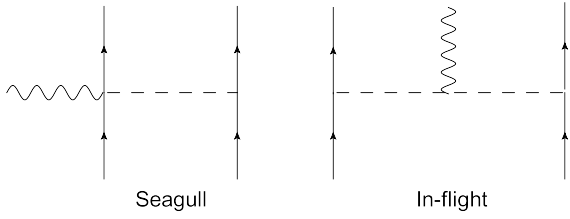}}
\caption{ Diagrams of the one-pion
exchange current: seagull (left) and in-flight (right).}
\label{fig1}
\end{figure}

Although there is
no explicit pion meson in RMF theory, it is possible to study the meson exchange current (MEC) corrections due to the virtual pion exchange between two
nucleons, which, according to Ref.~\cite{Morse1990}, are given by the two Feynman diagrams in Fig.~\ref{fig1}.

The one-pion exchange current contributions to magnetic moments are given by,
\begin{eqnarray}\label{magnetic moment-MEC}
    \mbox{\boldmath{$\mu$}}_{\mathrm{MEC}}
    &=& \frac{1}{2}\int d \br\,\br\times
    \langle g.s.|\hat\bj^{\mathrm{seagull}}(\br)
    +\hat\bj^{\mathrm{in\mbox{-}flight}}(\br)|g.s.\rangle,\nonumber\\
\end{eqnarray}
with the corresponding one-pion exchange currents
$\hat\bj^{\mathrm{seagull}}(\br)$ and
$\hat\bj^{\mathrm{in\mbox{-}flight}}(\br)$~\cite{Li2011b},
\begin{widetext}
\begin{subequations}
\begin{eqnarray}
    \hat\bj^{\mathrm{seagull}}(\br)
    &=&-\frac{8ef^2_{\pi}M}{m^2_\pi} \int d \svec x\,
        \bar{\psi}_p(\svec r) {\bm\gamma}\gamma_5\psi_n(\svec r)
         D_\pi(\svec r,\svec x)
    \bar{\psi}_n(\svec x)\frac{M^*}{M}\gamma_5\psi_p(\svec x),\\
    \hat\bj^{\mathrm{in\mbox{-}flight}}(\br)
    &=&-\frac{16ief^2_{\pi}M^2}{m_\pi^2} \int d\svec x d\svec y \bar{\psi}_p(\svec x)\frac{M^*}{M}\gamma_5\psi_n(\svec x)
        D_\pi(\svec x,\svec r){\bm\nabla}_{\svec r}
        D_\pi(\svec r,\svec y)\times\bar{\psi}_n(\svec y)\frac{M^*}{M}\gamma_5\psi_p(\svec
        y).
\end{eqnarray}
\end{subequations}
\end{widetext}
The pion propagator in r-space has the form
$D_\pi(\svec x,\svec r)=\dfrac{1}{4\pi}\dfrac{e^{-m_\pi|\svec
x-\svec r|}} {|\svec x-\svec r|}$.

\subsection{Configuration mixing}
The residual interaction, neglected in the mean field approximation, leads to configuration mixing, i.e., the coupling between the valence nucleon and particle-hole states in the core. This is also called core polarization. The configuration mixing corrections to the magnetic moment is treated approximately by Rayleigh-Schr\"{o}dinger perturbation theory.

\subsubsection{First-order corrections}

\begin{figure}
\centerline{
\includegraphics[width=8.5cm]{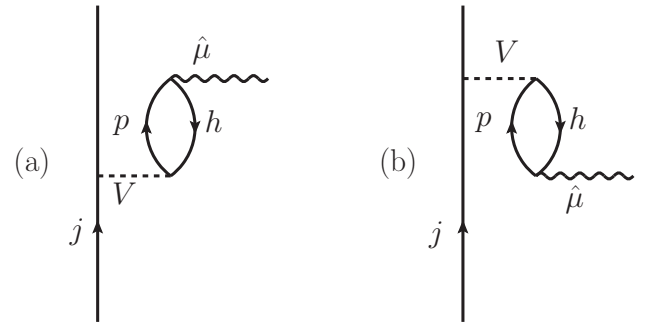}}
\caption{ Diagrams of first-order configuration mixing corrections to the magnetic moment. The external line represents the valence nucleon, and the intermediate
particle-hole pair represents an excited state of the core.}
\label{fig2}
\end{figure}

According to Rayleigh-Schr\"{o}dinger perturbation theory, the
first-order correction to the magnetic moments is determined as
\begin{eqnarray}\label{eq:1st}
  \delta\mu_{\mathrm{1st}}
  &=& \langle n|\hat{\mu}\frac{\hat{Q}}{E_n-\hat{H}_0}\hat{V}|n\rangle
  + \langle n|\hat{V}\frac{\hat{Q}}{E_n-\hat{H}_0}\hat{\mu}|n\rangle,\nonumber\\
\end{eqnarray}
where $\vert n\rangle$ and $E_n$ denote the unperturbed ground-state wavefunctions and corresponding energies. $\hat H_0$ and $\hat V$ are the operators of the mean-field Hamiltonian and the residual interaction respectively. $\hat{\mu}$ is the magnetic moment operator, and $\hat Q$ projects onto multi-particle and multi-hole configurations. The corresponding Feynman diagram is shown in Fig.~\ref{fig2}, where the wiggly lines represent the external field (here the magnetic moment operator), and the dashed lines denote the residual interaction. Solid lines with arrow upwards denote particle states (i.e., single particle orbits above the Fermi surface) and those with arrow downwards are hole states (i.e., single particle orbits in the Fermi sea).

For the magnetic moments of nuclei with a doubly closed shell core $\pm1$ nucleon, the formula for the first-order correction can be simplified as
\begin{widetext}
\begin{equation}\label{eq:1cp-bq}
    \delta\mu_{1\mathrm{st}}=\sum_{j_pj_hJ}
    \frac{2\langle j_h\|\bmu\|j_p\rangle}{\Delta E_j}
    (-1)^{j_h+j+J}\hat{j\,}^{-1}\sqrt{\frac{j}{j+1}}(2J+1)
    \left\{
      \begin{array}{ccc}
        j_h & j_p & 1 \\
        j & j & J \\
      \end{array}
    \right\}
    \langle jj_p;JM|V|jj_h;JM\rangle.
\end{equation}
\end{widetext}
$j$ denotes the valence nucleon state, $j_p$ and $j_h$ represent particle and hole states.
$\Delta E=\varepsilon_{j_p}-\varepsilon_{j_h}$ is the excitation energy of the one-particle-one-hole (1p-1h) excitation.
The selection rule $\Delta\ell=0$ of the non-relativistic magnetic moment operator allows only particles and holes as spin-orbit
partners, i.e., $j_p=\ell-\frac{1}{2}$ and $j_h=\ell+\frac{1}{2}$. All other diagrams vanish. It should be noted that the first-order configuration mixing
does not provide any contribution in nuclei with an $LS$ closed core $\pm1$ nucleon, because there are no spin-orbit
partners on both sides of the Fermi surface and therefore the magnetic-moment operator cannot couple to
magnetic resonances~\cite{Arima1987}.

\subsubsection{Second-order corrections}
\begin{figure}[h]
\centerline{
\includegraphics[width=8.5cm]{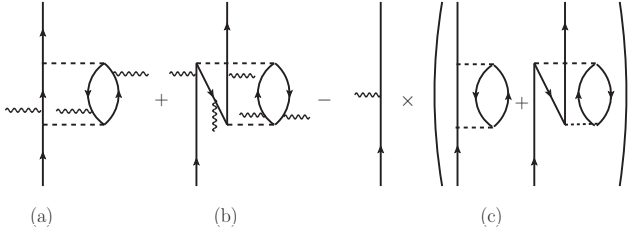}}
\caption{ Diagrams of second-order configuration mixing corrections to magnetic moment:
(a) 1p-1h mode, (b) 2p-2h mode and (c) wavefunction renormalization. Diagrams with more than one external wiggly
lines are an abbreviation for several separate diagrams where each of them has only one wiggly line at the indicated places.}
\label{fig3}
\end{figure}
As it is shown in Refs.~\cite{Shimizu1974,Arima1987} the second-order correction to the magnetic moments given by
\begin{eqnarray}\label{eq:sec}
  \delta\mu_{\mathrm{cm}}^{\mathrm{2nd}}
   &=& \langle n|\hat V\frac{\hat Q}{E_n-\hat H_0}\hat{\mu}\frac{\hat Q}{E_n-\hat H_0}\hat V|n\rangle\nonumber\\
   & & -\langle n|\hat{\mu}|n\rangle
        \langle
        n|\hat V\frac{\hat Q}{(E_n-\hat H_0)^2}\hat V|n\rangle,
\end{eqnarray}
where the second term comes from the renormalization of nuclear wave
function. In the second-order corrections we have to include one-particle-one-hole and two-particle-two-hole (2p-2h) contributions.
As shown in Fig.~\ref{fig3}, for a system with a doubly closed shell core, the second-order correction to the magnetic moment can be divided
into three terms~\cite{Arima1987}, the contributions of two-particle one-hole (2p-1h), of three-particle two-hole (3p-2h) configurations,
and of the wave function renormalization respectively.

\begin{figure}
 \centering
 \includegraphics[width=8.5cm]{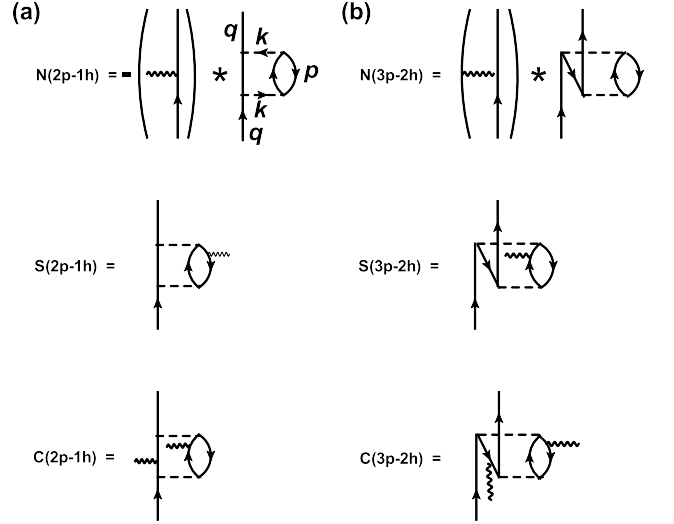}\\
 \caption{Diagrams representing second-order configuration mixing
corrections: (a) 2p-1h and (b) 3p-2h intermediate states.
For the notations N, S, and C, see the text for details.
\label{fig4}}
\end{figure}

As shown in Fig.~\ref{fig4}, we find for a system with a doubly closed core plus one particle the following second order corrections to the magnetic moment:
\begin{itemize}
  \item  N(2p-1h) and N(3p-2h): from the wave function renormalization.
  \item  S(2p-1h) and C(3p-2h): the external field operator acting on the hole line.
  \item  C(2p-1h) and S(3p-2h): the external field operator acting on the particle
  line.
\end{itemize}

and we obtain in this case
\begin{eqnarray}
    \delta\mu_{\mathrm{cm}}^{\mathrm{2nd}}
   &=& ~\mbox{N(2p-1h)} + \mbox{S(2p-1h)} + \mbox{C(2p-1h)}\nonumber\\
   & & + \mbox{N(3p-2h)} + \mbox{S(3p-2h)} + \mbox{C(3p-2h)}.
\end{eqnarray}
For a system of a doubly closed core minus one nucleon we have
\begin{eqnarray}
    \delta\mu_{\mathrm{cm}}^{\mathrm{2nd}}
   &=& ~\mbox{N(2h-1p)} + \mbox{S(2h-1p)} + \mbox{C(2h-1p)}\nonumber\\
   & & + \mbox{N(3h-2p)} + \mbox{S(3h-2p)} + \mbox{C(3h-2p)},
\end{eqnarray}
The detailed formulas for each term can be found in the appendix.

\section{Numerical details}
In this paper we study the magnetic moments of the nuclei $^{207}$Pb, $^{209}$Pb, $^{207}$Tl and $^{209}$Bi. We start on the mean field level with the magnetic moments derived from RMF theory including time-odd fields and add one-pion exchange currents and first- and second-order configuration mixing contributions. In the calculations, the relativistic point-coupling model with the density functional PC-PK1~\cite{Zhao2010} is applied.

Spherical RMF theory is solved in coordinate space, with a box size of $15\, \mathrm{fm}$ and a step size of $0.1\,\mathrm{fm}$. To calculate the corrections resulting from time-odd fields, triaxially deformed RMF calculations with time-odd fields are performed and each Dirac spinor is expanded in terms of a set of a three-dimensional harmonic
oscillator (HO) basis in Cartesian coordinates with 10 major shells~\cite{Koepf1989}. Pairing correlations in the vicinity of the doubly magic nucleus $^{208}$Pb are neglected.
The one-pion exchange current and configuration mixing corrections to the magnetic moments are calculated using spherical Dirac spinors of the nearby doubly closed shell nucleus $^{208}$Pb. The $\pi$-nucleon coupling constant is $f_\pi = 1$ and the pion mass $m_\pi = 138$ MeV.
The configuration mixing corrections depend on the configuration space and this will be discussed in the following applications.

\section{Results and discussion}
\subsection{First-order corrections}
In Table.~\ref{tab1} we give the first-order configuration mixing corrections to the magnetic moment of $^{209}$Bi. They are obtained
from relativistic calculations using the density functional PC-PK1~\cite{Zhao2010} and they are compared with the non-relativistic results using different interactions: Kallio-Kolltveit (KK)~\cite{Green1965}, Gillet~\cite{Gillet1964}, Kim-Rasmussen (KR)~\cite{Kim1963}, Brueckner~\cite{Brueckner1958}, Hamada-Johnston (HJ)~\cite{Hamada1962} potential, and the M3Y~\cite{Bertsch1977} interaction. The corresponding first-order corrections are taken from Refs.~\cite{Mavromatis1966,Mavromatis1967,Arima1972} respectively. For the relativistic calculations, we present both results with and without considering the residual interaction provided by pion exchange.

\begin{table*}
\tabcolsep=1.6pt
\centering
\caption{
\label{tab1}
First-order configuration mixing corrections to the magnetic moment of $^{209}$Bi obtained from relativistic calculations using the PC-PK1 effective interaction, in comparison with non-relativistic results using different interactions. In the relativistic calculations, both results with and without considering the residual interaction provided by pion are given.}
\small{
\begin{ruledtabular}
\begin{tabular}{ccrcrcccccrc}
  & \multicolumn{8}{c}{Non-rel.} && \multicolumn{2}{c}{Rel.}\\ \cline{2-9}\cline{11-12}
Interactions & KK & Gillet & KR. I & KR. II & Brueckner & HJ & ~Kuo~ & M3Y && PC-PK1  & PC-PK1
\\ \cline{3-7}\cline{11-12}
Ref.&\cite{Mavromatis1966}&\multicolumn{5}{c}{\cite{Blomqvist1965,Mavromatis1967}}&\cite{Arima1972}&\cite{Bertsch1977} && \multicolumn{1}{c}{without $\pi$}&\multicolumn{1}{c}{with $\pi$}\\\hline
$(1h_{\frac{9}{2}}1h_{\frac{11}{2}}^{-1})_{\pi}$
& 0.37 & 0.46 & 0.53 & 0.70 & 0.71 & 0.55 &  &0.43 &~~~&$-$0.11&0.19\\
$(1i_{\frac{11}{2}}1i_{\frac{13}{2}}^{-1})_{\nu}$
& 0.15 &$-$0.02 & 0.00 & $-$0.06 & 0.04 & 0.25 & &0.24 && 0.07 & 0.33\\
Total  & 0.52 & 0.43 & 0.53 & 0.64 & 0.75 & 0.80 & 0.79 &0.68 && $-0.04$ & $0.52$\\
\end{tabular}
\end{ruledtabular}}
\end{table*}

In non-relativistic calculations, only the excitations of spin-orbit partners can contribute to the first-order magnetic moment correction, because of selection rules imposed by the magnetic moment single-particle operator. In relativistic calculations this selection rule is only approximately valid. However, in the present relativistic calculations, only two particle-hole excitations can contribute to the first-order magnetic moment correction of $^{209}$Bi, i.e., the $(1h_{\frac{9}{2}}1h_{\frac{11}{2}}^{-1})_{\pi}$ and $(1i_{\frac{11}{2}}1i_{\frac{13}{2}}^{-1})_{\nu}$ excitations, and all other particle-hole excitations give a small and negligible contribution to its expectation value in first-order perturbation theory.

 \begin{figure}[h!]
 \centering
 \includegraphics[width=8.5cm]{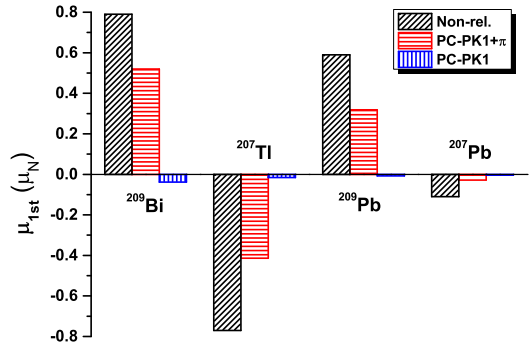}\vspace*{-1.0cm}
  \caption{(Color online) First-order configuration mixing corrections to magnetic moments of $^{209}$Bi, $^{207}$Tl, $^{209}$Pb and $^{207}$Pb obtained from relativistic calculations using PC-PK1 interaction, in comparison with the non-relativistic results obtained from Ref.~\cite{Arima1972}. In the relativistic calculations, the results with and without $\pi$ are given.}
 \label{fig5}
 \end{figure}

As shown in Table~\ref{tab1} the non-relativistic calculations give remarkable first-order corrections ($0.43\,\mu_N\sim0.80\,\mu_N$),
while the corresponding corrections given by relativistic calculations using the PC-PK1 effective interaction are very small ($-0.03\,\mu_N$)
and can be neglected. Only after the residual interaction provided by the pion is included, PC-PK1 gives significant corrections ($0.59\,\mu_N$)
that are consistent with non-relativistic results.

In order to further confirm the effects of the residual interaction provided by pion, we show in Fig.~\ref{fig5} the first-order configuration mixing corrections to the magnetic moments of $^{207}$Pb, $^{209}$Pb, $^{207}$Tl and $^{209}$Bi obtained from relativistic calculations using the PC-PK1 interaction. They are compared with non-relativistic results obtained from Ref.~\cite{Arima1972}. In Fig.~\ref{fig5} we also see that without the residual interaction provided by pion, the relativistic calculations give negligible first-order corrections to the magnetic moments of all four nuclei. If the residual interaction provided by pion is included, relativistic calculations are in reasonable agreement with non-relativistic results for all present nuclei.

Considering Eq.~(\ref{eq:1cp-bq}), we find that the magnetic moment operator, the excitation energy, and the interactions can all lead to differences between relativistic
and non-relativistic results.
It is well known that the effective mass is relatively small in self-consistent calculations based on density functional theory. This leads to an increased gap at the Fermi surface in the single particle spectrum and to larger ph-energies. This is a deficiency of conventional density functional theory
based on the mean field approximation with energy independent self energies. Taking into account the energy dependence of the self energy in the framework of couplings to
low-lying collective surface modes considerably larger effective masses and smaller energy gaps have been found in the literature~\cite{Ring1973,Bernard1980,Litvinova2006,Ring2009},
which are closer to the experimental values. Such calculations go, obviously, beyond the scope of the present investigations and
therefore we use the experimental single particle energies rather than the self-consistent RMF single particle energies in the intermediate states for a relativistic estimation.
The experimental excitation energies
used in these calculations are $\Delta E_p=\varepsilon(1h_{\frac{9}{2}})-\varepsilon(1h_{\frac{11}{2}})=5.6$~MeV and $\Delta~E_n=\varepsilon(1i_{\frac{11}{2}})-\varepsilon(1i_{\frac{13}{2}})=5.86$~MeV. Since the non-relativistic results are obtained with experimental energy splittings, it is found that by adopting experimental excitation
energy, the relativistic calculations give almost
the same first-order corrections as the results by adopting self-consistent RMF single particle energies shown in Fig.~\ref{fig5}.
Although there is some difference between the matrix elements of the relativistic and the non-relativistic magnetic moment operator, this causes not much difference in
the first-order corrections in Fig.~\ref{fig5}. Therefore, the difference between relativistic and non-relativistic results are mainly due to interactions, and the
residual interaction provided by pion plays an important role in the relativistic descriptions of nuclear magnetic moments. Therefore it will be included in the following
calculations of second-order corrections.

\subsection{Second-order corrections}

\begin{figure}[h!]
 \centering
 \includegraphics[width=8.5cm]{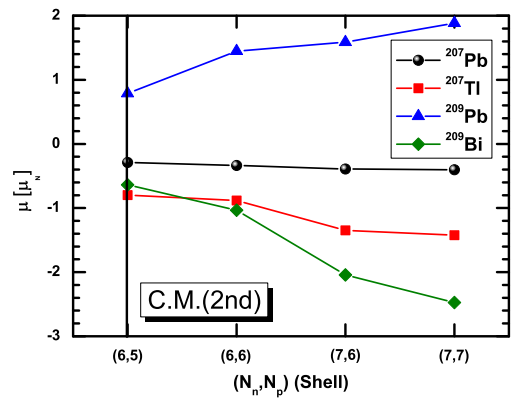}
 \caption{(Color online) The second-order configuration mixing corrections to the magnetic moments for major shell
 truncations, i.e., the sum in the intermediate states is restricted to single particle energies in major shells with the quantum numbers ($N_n,N_p$).}
 \label{fig6}
\end{figure}

The theoretical analysis for the Feynman diagrams of second-order configuration mixing shows that the corrections to the magnetic moments are diverging as the
configuration space is increased. This is discussed in details in the appendix.
In order to investigate the relationship between second-order corrections and corresponding configuration space, and also to choose the appropriate truncation,
the second-order configuration mixing corrections to the magnetic moments of $^{207}$Pb, $^{209}$Pb, $^{207}$Tl and $^{209}$Bi are given in Fig.~\ref{fig6} for major shell truncations~\cite{Shimizu1974}, i.e., the configuration space for the calculations of the single-particle levels in the intermediate states is restricted by the major
shells with the quantum numbers $2(n-1)+\ell\leq N_n$ for neutrons and $2(n-1)+\ell\leq N_p$ for protons correspondingly.

The second-order corrections of the four nuclei become numerically larger with increasing of the configuration space. Only the corrections of $^{207}$Pb are relatively small and change only slightly. The present second-order corrections of the four nuclei strongly depend on the orbit of corresponding valence nucleon. In addition, the truncations which give the best descriptions of present four nuclei are labelled with vertical line in Fig.~\ref{fig6} ($N_n=6,N_p=5$, i.e., one major shell above the closed shells with the neutron number $N=126$ and proton number $Z=82$). This truncation gives the smallest relative deviations $6.1\%$ between the relativistic calculations and experimental data, where the relativistic calculations include the magnetic moments from RMF theory, contributions from the corresponding time-odd fields, the one-pion exchange currents and the first-order as well as second-order corrections. In order to avoid divergent results in perturbation theory a truncation is necessary. The present truncation of one major shell above the closed core is certainly only a first attempt. It is however reasonable, because the important physics takes place in this space. It is also clear that all the non-relativistic investigations of higher order configurations mixing in the literature are based on finite configuration space. Investigations to treat such diverging diagrams in theories based on density functional theory and going beyond the mean field concept are in their infancy~\cite{Litvinova2007,Moghrabi2012,Moghrabi2012a}.

\begin{figure*}
\centering
\includegraphics[width=7.5cm]{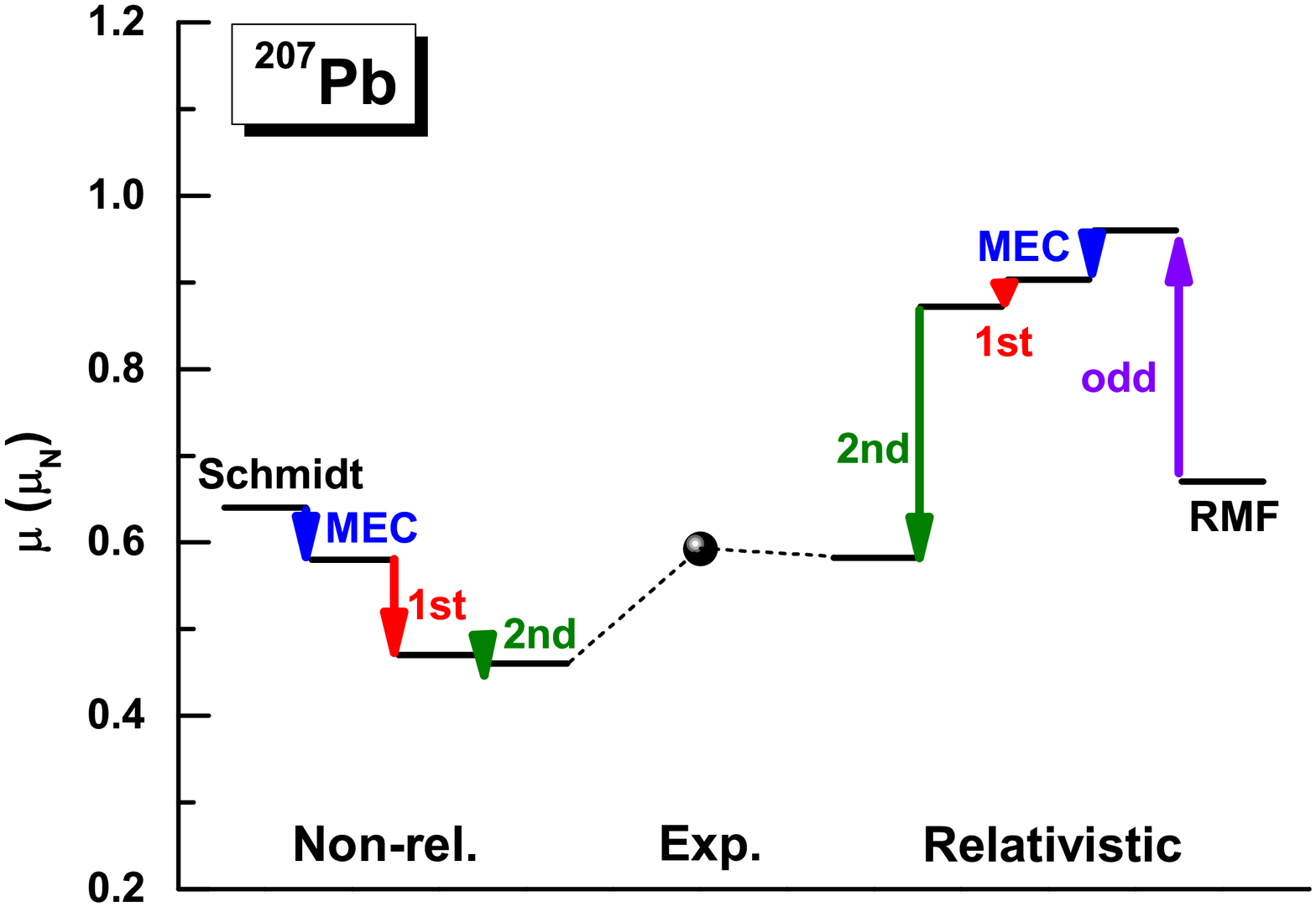} \hspace*{-0.8cm}
\includegraphics[width=7.5cm]{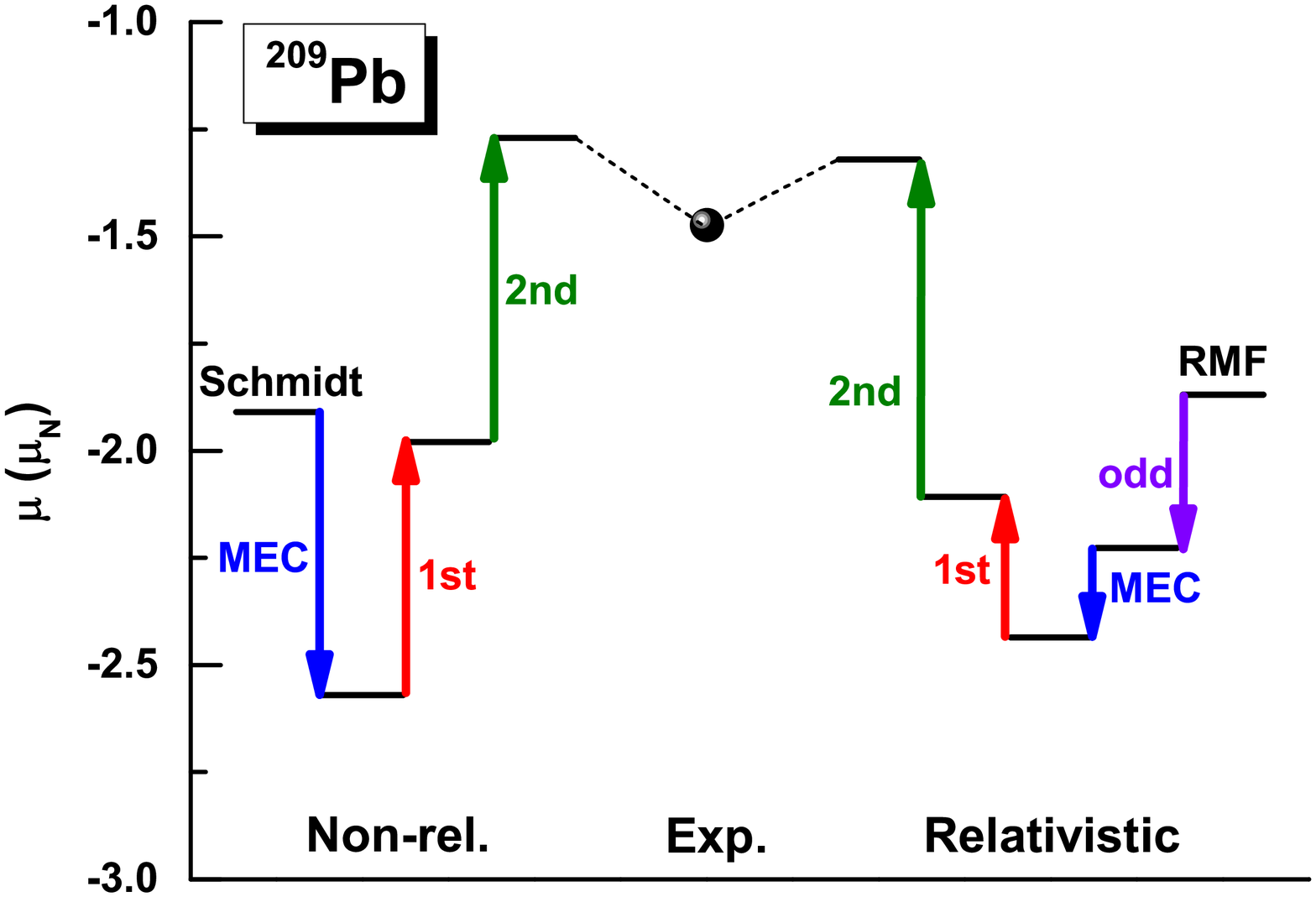} \vspace*{-1.0cm}\\
\includegraphics[width=7.5cm]{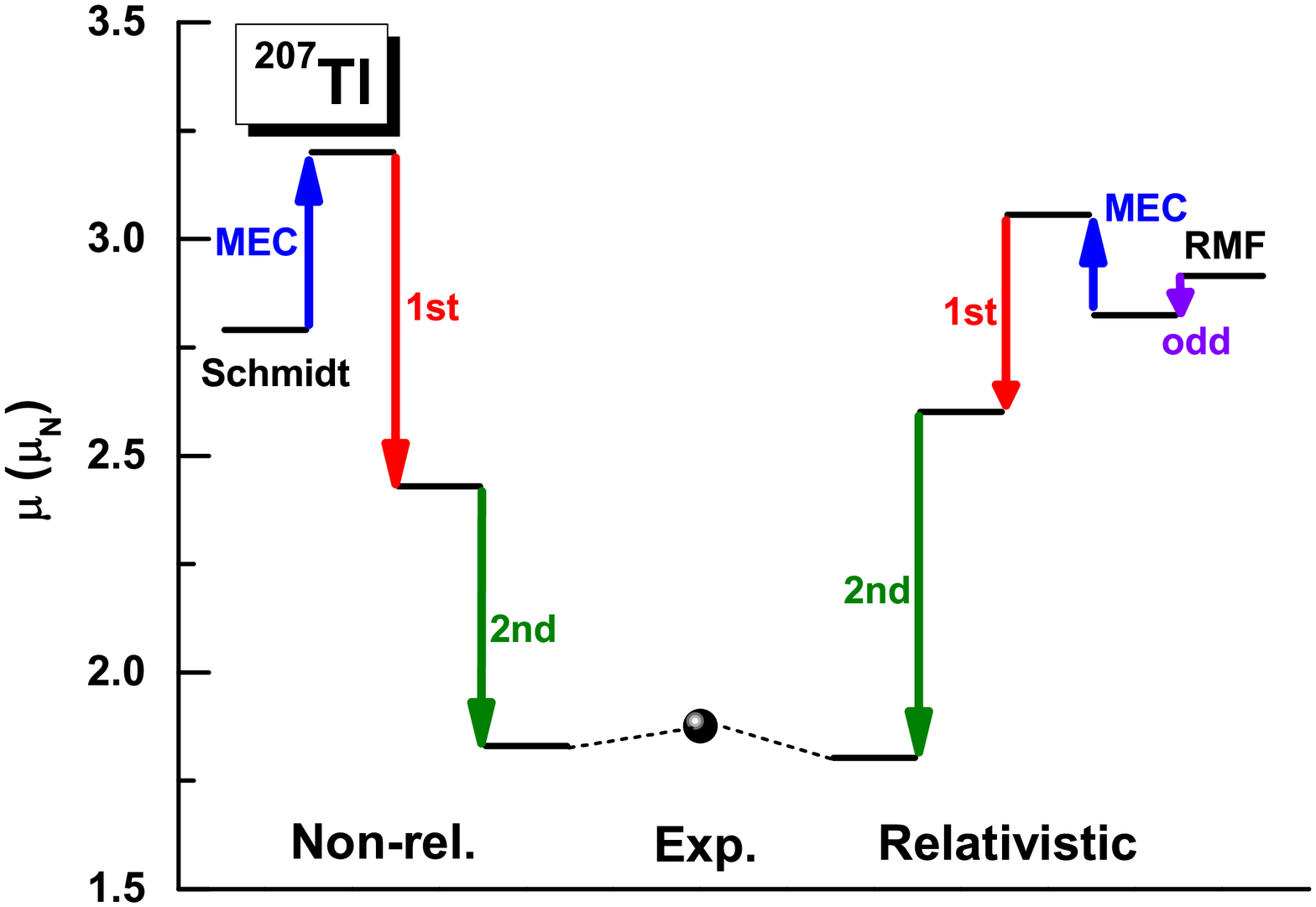} \hspace*{-0.8cm}
\includegraphics[width=7.5cm]{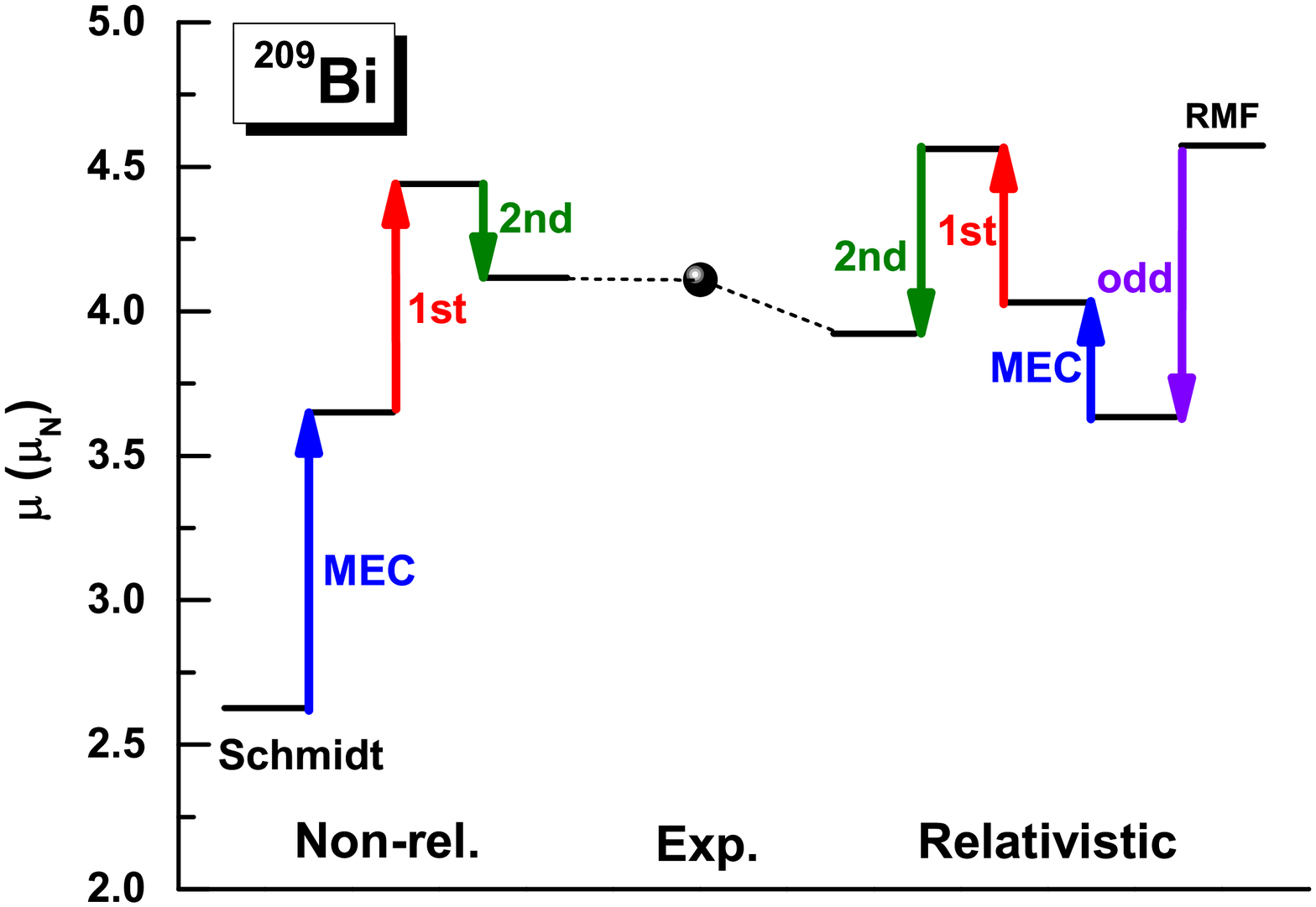} \vspace*{-0.5cm}
\caption{(Color online) Magnetic moments of the nuclei $^{207}$Pb, $^{209}$Pb, $^{207}$Tl and $^{209}$Bi, obtained from relativistic calculations using the PC-PK1 interaction with time odd contribution, meson exchange currents, first- and second-order corrections, in comparison with data (solid circle) and the corresponding non-relativistic results from Ref.~\cite{Arima1987}.}
\label{fig7}
\end{figure*}

\subsection{The magnetic moments of nuclei near $^{208}$Pb}

Fig.~\ref{fig7} shows the final results for the magnetic moments of $^{207}$Pb, $^{209}$Pb, $^{207}$Tl and $^{209}$Bi. They are obtained from RMF theory using the density functional PC-PK1 and corresponding corrections are added: time-odd fields (labeled as odd), meson exchange currents (labeled as MEC), first- (labeled as 1st) and second-order (label as 2nd) configuration mixing. The truncation is $2(n-1)+\ell\leq 6$ for the neutron and $2(n-1)+\ell\leq 5$ for the proton levels. These relativistic results are compared with data (labeled as solid circle) and non-relativistic results from Ref.~\cite{Arima1987}. The magnetic moments obtained from spherical RMF theory are labeled as RMF. The differences between magnetic moments of triaxial deformed theory with time-odd fields and magnetic moments of spherical theory represent the corrections due to time-odd fields. In the relativistic calculations, MEC only contains the one-pion exchange current correction, while the MEC in non-relativistic calculations includes the one-pion exchange current, the $\Delta$ isobar current and the crossing term between MEC and first-order configuration mixing. For first- and second-order corrections in the relativistic calculation, the residual interaction provided by the poin is included.

It is seen from Fig.~\ref{fig7} that in the relativistic calculations the magnetic moments of all four nuclei are considerably improved by including first-order corrections,
MEC and second-order corrections, and they are now in agreement with non-relativistic results. The magnetic moment of $^{207}$Pb is excellently reproduced by relativistic calculations, while the corresponding deviation from data is $0.01\,\mu_N$, and much better than the non-relativistic deviation $0.05\,\mu_N$.
For $^{209}$Pb, the deviations from data are about $0.15\,\mu_N$ and $0.2\,\mu_N$ respectively for relativistic and non-relativistic results. For the magnetic moment
of $^{207}$Tl, both the non-relativistic description and the relativistic description are very good, as the corresponding two deviations from data are less than $0.1\,\mu_N$.
The magnetic moment of $^{209}$Bi is also well reproduced by relativistic and non-relativistic calculations, and the relative deviations from data for both calculations are less than $5\%$. On the whole, the relative deviation $\sigma_r$ of the present four nuclei in the relativistic calculation is $6.1\%$, better than the corresponding non-relativistic results $13.2\%$~\cite{Arima1987}.


It is obvious that the first-order, MEC and second-order corrections given by relativistic calculations have the same sign and order of magnitude as the corresponding corrections given by non-relativistic calculations. This further shows that the present relativistic calculations are reasonable, including the appropriate treatment of truncation in second-order corrections.

\section{Summary}
In summary, using the relativistic point-coupling model based on the density functional PC-PK1 the magnetic moments of the nuclei $^{209}$Pb, $^{207}$Pb, $^{209}$Bi and $^{207}$Tl with a $jj$ closed-shell core $^{208}$Pb are studied, based on the magnetic moments from RMF theory and corresponding the time-odd fields, one-pion exchange current, first- and second-order corrections. It is found that the second order diagrams diverge, if the sum over the intermediate single particle states is carried out to infinity. Therefore a reasonable cut off is introduced in this sum and only states in the first major shell above the Fermi surface ($N_n=6$ for neutrons and $N_p=5$ for protons) are taken into account. The data are well reproduced by the present relativistic calculations. They are compared with corresponding non-relativistic results from the literature. In general, the relative deviation of $6.1\%$ from experiment for the four nuclei obtained in relativistic calculations is better than the corresponding non-relativistic results $13.2\%$. It is found that the pion is important to describe magnetic moments by means of one-pion exchange current and by the residual interaction provided by pion exchange.



Of course there are still many important open questions. It remains to consider the contribution from the Dirac sea, the crossing terms between MEC and configuration mixing, the influence of higher order diagrams in RPA type configuration mixing calculations~\cite{Bauer1973}, and coupling to the $\Delta$ isobar current~\cite{Rho1974,Oset1979,Knupfer1980}.
Of course, it will be also interesting to study the influence of other successful covariant density functionals on the market, in particular those bases on relativistic Hartree-Fock theory~\cite{Long2006}, where the pion and the resulting tensor forces can be included in a self-consistent way. Work in this direction is progress.

\begin{acknowledgments}
We would like to thank Akito Arima, Haozhao Liang and Jiangming Yao for discussions and collaboration. This work is partly supported by the Major State 973 Program 2013CB834400, the NSFC (Grant Nos. 11175002 and 11205068) and CPSC (Grant Nos. 2012M520100 and 2012M520667) and by the DFG cluster of excellence \textquotedblleft Origin and Structure of the Universe\textquotedblright\ (www.universe-cluster.de).

\end{acknowledgments}

\section{Appendix}
\subsection{Second-order corrections}
In the second-order correction, both 1p-1h and 2p-2h excitation modes can be divided into three parts, N, S and C respectively~\cite{Shimizu1974,Arima1987},
\begin{subequations}
\begin{eqnarray}
  \mathrm{N} &=& \mp\langle j|\hat{\mu}|j\rangle\langle j|V\frac{\hat{Q}}{(E_j^{(0)}-\hat{H}_0)^2}V|j\rangle,\label{eq:2cp-N}\\
  \mathrm{S} &=& \langle j|V\frac{\hat{Q}}{E_j^{(0)}-\hat{H}_0}\hat{\mu}\frac{Q}{E_j^{(0)}-\hat{H}_0}V|j\rangle,  \\
  \mathrm{C} &=& \langle j|V\frac{\hat{Q}}{E_j^{(0)}-\hat{H}_0}\hat{\mu}\frac{Q}{E_j^{(0)}-\hat{H}_0}V|j\rangle.
\end{eqnarray}
\end{subequations}
In Eq.~(\ref{eq:2cp-N}), the minus sign $-$ is for 1p-1h mode, and + is for 2p-2h mode. For the nuclei of doubly
closed shells plus one nucleon, the corresponding second-order correction can be written as
\begin{subequations}
\begin{eqnarray}\label{eq:2cp-2p1h-N}
  \mbox{N(2p-1h)}
   &=& -\langle j\|\bmu\|j\rangle\sum_{j_1j_2j_h,J}\frac{\hat{J}^2\hat{j}^{-2}}{\Delta E^2_j}\langle jj_h,J|V|j_1j_2,J\rangle^2,\nonumber\\
\end{eqnarray}
\begin{eqnarray}\label{eq:2cp-2p1h-S}
  \mbox{S(2p-1h)}
    &=& -\sum_{j_{h}j_{h'}}\sum_{j_1j_2,J}
        (-1)^{j_{h'}+J+j}\hat{J}^2
         \left\{
           \begin{array}{ccc}
             j_h & j_{h'} & 1 \\
             j & j & J \\
           \end{array}
         \right\}\nonumber\\
    & &\times\,\frac{1}{\Delta E\Delta E'}\langle j_{h'}\|\bmu\|j_h\rangle\langle jj_h,J|V|j_1j_2,J\rangle\nonumber\\
    & &\times\,\langle  j_1j_2,J|V|jj_{h'},J\rangle,
\end{eqnarray}
\begin{eqnarray}\label{eq:2cp-2p1h-C}
  \mbox{C(2p-1h)}
    &=& \sum_{j_1j_2,j'_{1}j_{h}}\sum_{JJ'}(-1)^{j_1+j_2+j_h+j}
         2\hat{J}^2\hat{J'}^2\langle j_1\|\bmu\|j'_1\rangle\nonumber\\
    & & \times\,
         \left\{
           \begin{array}{ccc}
             J & 1 & J' \\
             j'_1 & j_2 & j_1 \\
           \end{array}
         \right\}
       \left\{
           \begin{array}{ccc}
             J' & 1 & J \\
             j & j_h & j \\
           \end{array}
         \right\}\frac{1}{\Delta E\Delta E'}\nonumber  \\
  & &\times\,\langle jj_h,J|V|j_1j_2,J\rangle\langle
  j'_1j_2,J'|V|jj_{h},J'\rangle,\nonumber  \\
\end{eqnarray}

\begin{eqnarray}\label{eq:2cp-3p2h-N}
  \mbox{N(3p-2h)}
   &=& -\langle j\|\bmu\|j\rangle\sum_{j_{h_1}j_{h_2}}^{j_1,J}\frac{\hat{J}^2\hat{j}^{-2}}{\Delta E^2_j}
       \langle jj_1,J|V|j_{h_1}j_{h_2},J\rangle^2,\nonumber\\
\end{eqnarray}

\begin{eqnarray}\label{eq:2cp-3p2h-S}
  \mbox{S(3p-2h)}
    &=& -\sum_{j_{h_1}j_{h_2}}\sum_{j_1j_2,J}
        \frac{(-1)^{j_1+J+j}\hat{J}^2}{\Delta E\Delta E'} \left\{
           \begin{array}{ccc}
             j_{1} & j_{2} & 1 \\
             j & j & J \\
           \end{array}
         \right\}\nonumber\\
    & & \times\, \langle j_1\|\bmu\|j_{2}\rangle\langle jj_1,J|V|j_{h_1}j_{h_2},J\rangle\nonumber\\
    & & \times\, \langle  j_{h_1}j_{h_2},J|V|jj_{2},J\rangle,
\end{eqnarray}

\begin{eqnarray}\label{eq:2cp-3p2h-C}
  \mbox{C(3p-2h)}
    &=& \sum_{j_{h_1}j_{h_2}j_{h'_1}}\sum_{j_{1}JJ'}(-1)^{j_{h_1}+j_{h_2}+j_1+j}
         \langle j_{h_1}\|\bmu\|j_{h'_1}\rangle\nonumber\\
    & & \times\,\frac{2\hat{J}^2\hat{J'}^2}{\Delta E\Delta E'}
         \left\{
           \begin{array}{ccc}
             J & 1 & J' \\
             j_{h'_1} & j_{h_2} & j_{h_1} \\
           \end{array}
         \right\}
       \left\{
           \begin{array}{ccc}
             J' & 1 & J \\
             j & j_1 & j \\
           \end{array}
         \right\}\nonumber  \\
  & &\times\, \langle jj_1,J|V|j_{h_1}j_{h_2},J\rangle\langle
  j_{h'_1}j_{h_2},J'|V|jj_{1},J'\rangle.\nonumber\\
\end{eqnarray}
\end{subequations}
In all formulas, a factor $\langle jj10|jj\rangle/\sqrt{2j+1}$ should be accompanied. $\Delta E$ is the excitation energy of the 2p-1h (3p-2h) immediate states. For nuclei with one hole, the formulas for $N(\mathrm{2h-1p})$ etc. are simply given by interchanging the indices $p$ and $h$.

Some Feynman diagrams~\cite{Mattuck1992} for second-order corrections are not convergent. This can easily be seen in momentum space. Taking the term N(2p-1h) as an example, the Feynman diagram (shown in Fig.~\ref{fig4}) corresponds the following integration,
\begin{equation}\label{eq:N-2p-1h}
       \mathrm{N(2p-1h)}
       =-\langle j||~\hat{\mu}~||j\rangle\int_{\Gamma}\frac{d^3K}{(2\pi)^3}\frac{d^3P}{(2\pi)^3}|V_{K}|^2\frac{1}{\Delta E^2},
\end{equation}
where the integration space $\Gamma$ includes $|P|<k_{F}$, $|P+K|>k_{F}$, and $|q+K|>k_{F}$. $q,~K,~P$ denotes the momentum of the valence nucleon, the exchange momentum between the valence nucleon and the 1p-1h bubble, and the momentum of hole state, respectively. In the small or large momentum transform limit ($K\rightarrow0$ or $K\rightarrow\infty$), the excited energy of the intermediate state $\Delta E=\varepsilon_{q-K}+\varepsilon_{P+K}-\varepsilon_{P}-\varepsilon_{q}\sim K$. In the point-coupling model, $V_{K}$ is a constant and therefore
\begin{equation}
       \mathrm{N(2p-1h)}\sim\int d^3K\frac{1}{K^2}\cdots.
\end{equation}
Since $|q+K|>k_{F}$, $K$ can be infinite. Therefore, the N(2p-1h) term diverges.

After a similar analysis of the other terms, we found that for nucleus with core plus one nucleon (hole), the terms with 2p-1h~(3h-2p) intermediate states are not convergent either. The easiest way to deal with this divergence is to introduce an appropriate truncation in the integration space.

\subsection{The residual interaction in the relativistic point-coupling model}
As noted in Eq.~(\ref{eq:res.int.}), the relativistic residual
interaction is determined by the second derivative of the energy density
functional with respect to the density matrix. It can be expressed as $V^i_{\alpha\beta\alpha'\beta'}=\langle\alpha\beta|V^i|\alpha'\beta'\rangle$, with
\begin{eqnarray}
  V^S
   &=&
   \gamma_0(1)(\alpha_S+2\beta_S\rho_S+3\gamma_S\rho_S^2+\delta_S\Delta)_1
           \,\gamma_0(2)\nonumber\\
   & &\times\,\delta(\bm r_1-\bm r_2),
\end{eqnarray}
\begin{eqnarray}
  V^{V}
   &=&
   \{[\alpha_V+3\gamma_V\rho_V^2+\delta_V\Delta]_1-\balp(1)[\alpha_V+\gamma_V\rho_V^2\nonumber\\
   & &+\delta_V\Delta]_1\balp(2)\}\times\,
   \delta(\bm r_1-\bm r_2),
\end{eqnarray}
\begin{eqnarray}
  V^{TV}
   &=&
   [\gamma_0\gamma_\mu(\alpha_{TV}+\delta_{TV}\Delta)\vec{\tau}]_1\,
   [\gamma_0\gamma^\mu\vec{\tau}]_2\,\delta(\bm r_1-\bm r_2), \nonumber\\
\end{eqnarray}
\begin{equation}
    V^C= \frac{e^2}{4}[ \gamma_0\gamma_\mu(1-\tau_3)]_1 [
\gamma_0\gamma^\mu(1-\tau_3)]_2,
\end{equation}
which are for isoscalar, vector, isovector-vector and coulomb interaction, respectively.

With the pion-nucleon Lagrangian density in Eq.~(\ref{Eq:lag-PN}), the corresponding interaction reads
\begin{eqnarray}
  V^\pi
    = -[\frac{f_\pi}{m_\pi}\vec{\tau}\gamma_0\gamma_5\gamma^k\partial_k]_1\cdot
         [\frac{f_\pi}{m_\pi}\vec{\tau}\gamma_0\gamma_5\gamma^l\partial_l]_2D_\pi(1,2).
\end{eqnarray}

In order to cancel the contact interaction coming from the pion pseudovector coupling, a zero-range pionic counterterm is included, which reads
\begin{eqnarray}
    V^{\pi\delta}
    & = &
    \frac{1}{3}[\frac{f_\pi}{m_\pi}\vec{\tau}\gamma_0\gamma_5\bm\gamma]_1\cdot
    [\frac{f_\pi}{m_\pi}\vec{\tau}\gamma_0\gamma_5\bm\gamma]_2\delta(\bm r_1-\bm r_2).\nonumber\\
\end{eqnarray}


\begin{thebibliography}{50}
\expandafter\ifx\csname natexlab\endcsname\relax\def\natexlab#1{#1}\fi
\expandafter\ifx\csname bibnamefont\endcsname\relax
  \def\bibnamefont#1{#1}\fi
\expandafter\ifx\csname bibfnamefont\endcsname\relax
  \def\bibfnamefont#1{#1}\fi
\expandafter\ifx\csname citenamefont\endcsname\relax
  \def\citenamefont#1{#1}\fi
\expandafter\ifx\csname url\endcsname\relax
  \def\url#1{\texttt{#1}}\fi
\expandafter\ifx\csname urlprefix\endcsname\relax\def\urlprefix{URL }\fi
\providecommand{\bibinfo}[2]{#2}
\providecommand{\eprint}[2][]{\url{#2}}



\bibitem{Blin-Stoyle1956}
\bibinfo{author}{\bibfnamefont{R.~J.} \bibnamefont{Blin-Stoyle}},
  \bibinfo{journal}{Rev. Mod. Phys.} \textbf{\bibinfo{volume}{28}},
  \bibinfo{pages}{75} (\bibinfo{year}{1956}).

\bibitem{Arima1984}
\bibinfo{author}{\bibfnamefont{A.}~\bibnamefont{Arima}},
  \bibinfo{journal}{Prog. Part. Nucl. Phys.} \textbf{\bibinfo{volume}{11}},
  \bibinfo{pages}{53} (\bibinfo{year}{1984}).

\bibitem{Castel1990}
\bibinfo{author}{\bibfnamefont{B.}~\bibnamefont{Castel}} \bibnamefont{and}
  \bibinfo{author}{\bibfnamefont{I.~S.} \bibnamefont{Towner}},
  \emph{\bibinfo{title}{Modern Theories of Nuclear Moments}}
  (\bibinfo{publisher}{Clarendon Press on Oxford}, \bibinfo{year}{1990}).


\bibitem{Talmi2005a}
\bibinfo{author}{\bibfnamefont{I.}~\bibnamefont{Talmi}}, \bibinfo{journal}{Int. J.
  Mod. Phys. E} \textbf{\bibinfo{volume}{14}}, \bibinfo{pages}{821}
  (\bibinfo{year}{2005}).



\bibitem{Blin-Stoyle1953}
\bibinfo{author}{\bibfnamefont{R.~J.} \bibnamefont{Blin-Stoyle}},
  \bibinfo{journal}{Proc. Phys. Soc. A} \textbf{\bibinfo{volume}{66}},
  \bibinfo{pages}{1158} (\bibinfo{year}{1953}).

\bibitem{Schmidt1937}
\bibinfo{author}{\bibfnamefont{T.}~\bibnamefont{Schmidt}}, \bibinfo{journal}{Z.
  Phys.} \textbf{\bibinfo{volume}{106}}, \bibinfo{pages}{358}
  (\bibinfo{year}{1937}).

\bibitem{Towner1987}
\bibinfo{author}{\bibfnamefont{I.~S.} \bibnamefont{Towner}},
  \bibinfo{journal}{Phys. Rep.} \textbf{\bibinfo{volume}{155}},
  \bibinfo{pages}{263} (\bibinfo{year}{1987}).

\bibitem{Arima1987}
\bibinfo{author}{\bibfnamefont{A.}~\bibnamefont{Arima}},
  \bibinfo{author}{\bibfnamefont{K.}~\bibnamefont{Shimizu}},
  \bibinfo{author}{\bibfnamefont{W.}~\bibnamefont{Bentz}}, \bibnamefont{and}
  \bibinfo{author}{\bibfnamefont{H.}~\bibnamefont{Hyuga}},
  \bibinfo{journal}{Adv. Nucl. Phys.} \textbf{\bibinfo{volume}{18}},
  \bibinfo{pages}{1} (\bibinfo{year}{1987}).

\bibitem{Arima2011}
\bibinfo{author}{\bibfnamefont{A.}~\bibnamefont{Arima}},
  \bibinfo{journal}{Sci. China Phys. Mech. Astron.} \textbf{\bibinfo{volume}{54}},
  \bibinfo{pages}{188} (\bibinfo{year}{2011}).

\bibitem{Arima1954}
\bibinfo{author}{\bibfnamefont{A.}~\bibnamefont{Arima}} \bibnamefont{and}
  \bibinfo{author}{\bibfnamefont{H.}~\bibnamefont{Horie}},
  \bibinfo{journal}{Prog. Theor. Phys.} \textbf{\bibinfo{volume}{11}},
  \bibinfo{pages}{509} (\bibinfo{year}{1954}{\natexlab{a}}).

\bibitem{Talmi2005}
\bibinfo{author}{\bibfnamefont{I.}~\bibnamefont{Talmi}}, \bibinfo{journal}{J.
  Phys. Conf.} \textbf{\bibinfo{volume}{20}}, \bibinfo{pages}{28}
  (\bibinfo{year}{2005}).

\bibitem{Arima1954b}
\bibinfo{author}{\bibfnamefont{A.}~\bibnamefont{Arima}} \bibnamefont{and}
  \bibinfo{author}{\bibfnamefont{H.}~\bibnamefont{Horie}},
  \bibinfo{journal}{Prog. Theor. Phys.} \textbf{\bibinfo{volume}{12}},
  \bibinfo{pages}{623} (\bibinfo{year}{1954}{\natexlab{b}}).

\bibitem{Hiroshi1958}
\bibinfo{author}{\bibfnamefont{N.}~\bibnamefont{Hiroshi}},
  \bibinfo{author}{\bibfnamefont{A.}~\bibnamefont{Arima}}, \bibnamefont{and}
  \bibinfo{author}{\bibfnamefont{H.}~\bibnamefont{Horie}},
  \bibinfo{journal}{Prog. Theor. Phys. Supplement}
  \textbf{\bibinfo{volume}{8}}, \bibinfo{pages}{33} (\bibinfo{year}{1958}).

\bibitem{Miyazawa1951}
\bibinfo{author}{\bibfnamefont{H.}~\bibnamefont{Miyazawa}},
  \bibinfo{journal}{Prog. Theor. Phys.} \textbf{\bibinfo{volume}{6}},
  \bibinfo{pages}{801} (\bibinfo{year}{1951}).

\bibitem{Villars1952}
\bibinfo{author}{\bibfnamefont{F.}~\bibnamefont{Villars}},
  \bibinfo{journal}{Phys. Rev.} \textbf{\bibinfo{volume}{86}},
  \bibinfo{pages}{476} (\bibinfo{year}{1952}).

\bibitem{Chemtob1969}
\bibinfo{author}{\bibfnamefont{M.}~\bibnamefont{Chemtob}},
  \bibinfo{journal}{Nucl. Phys. A} \textbf{\bibinfo{volume}{123}},
  \bibinfo{pages}{449} (\bibinfo{year}{1969}).

\bibitem{Hyuga1980}
\bibinfo{author}{\bibfnamefont{H.}~\bibnamefont{Hyuga}},
  \bibinfo{author}{\bibfnamefont{A.}~\bibnamefont{Arima}}, \bibnamefont{and}
  \bibinfo{author}{\bibfnamefont{K.}~\bibnamefont{Shimizu}},
  \bibinfo{journal}{Nucl. Phys. A} \textbf{\bibinfo{volume}{336}},
  \bibinfo{pages}{363} (\bibinfo{year}{1980}).

\bibitem{Ichimura1965}
\bibinfo{author}{\bibfnamefont{M.}~\bibnamefont{Ichimura}} \bibnamefont{and}
  \bibinfo{author}{\bibfnamefont{K.} \bibnamefont{Yazaki}},
  \bibinfo{journal}{Nucl. Phys.} \textbf{\bibinfo{volume}{63}},
  \bibinfo{pages}{401} (\bibinfo{year}{1965}).

\bibitem{Mavromatis1966}
\bibinfo{author}{\bibfnamefont{H.~A.} \bibnamefont{Mavromatis}},
  \bibinfo{author}{\bibfnamefont{L.}~\bibnamefont{Zamick}}, \bibnamefont{and}
  \bibinfo{author}{\bibfnamefont{G.~E.} \bibnamefont{Brown}},
  \bibinfo{journal}{Nucl. Phys.} \textbf{\bibinfo{volume}{80}},
  \bibinfo{pages}{545} (\bibinfo{year}{1966}).


\bibitem{Mavromatis1967}
\bibinfo{author}{\bibfnamefont{H.~A.} \bibnamefont{Mavromatis}}
  \bibnamefont{and} \bibinfo{author}{\bibfnamefont{L.}~\bibnamefont{Zamick}},
  \bibinfo{journal}{Nucl. Phys. A} \textbf{\bibinfo{volume}{104}},
  \bibinfo{pages}{17} (\bibinfo{year}{1967}).

\bibitem{Shimizu1974}
\bibinfo{author}{\bibfnamefont{K.}~\bibnamefont{Shimizu}},
  \bibinfo{author}{\bibfnamefont{M.}~\bibnamefont{Ichimura}}, \bibnamefont{and}
  \bibinfo{author}{\bibfnamefont{A.}~\bibnamefont{Arima}},
  \bibinfo{journal}{Nucl. Phys. A} \textbf{\bibinfo{volume}{226}},
  \bibinfo{pages}{282} (\bibinfo{year}{1974}).

\bibitem{Towner1983}
\bibinfo{author}{\bibfnamefont{I.~S.} \bibnamefont{Towner}} \bibnamefont{and}
  \bibinfo{author}{\bibfnamefont{F.~C.} \bibnamefont{Khanna}},
  \bibinfo{journal}{Nucl. Phys. A} \textbf{\bibinfo{volume}{399}},
  \bibinfo{pages}{334} (\bibinfo{year}{1983}).


\bibitem{Reinhard1989}
\bibinfo{author}{\bibfnamefont{P. G.}~\bibnamefont{Reinhard}}, \bibinfo{journal}{Rep.
Prog. Phys.} \textbf{\bibinfo{volume}{52}}, \bibinfo{pages}{439}
  (\bibinfo{year}{1996}).

\bibitem{Ring1996}
\bibinfo{author}{\bibfnamefont{P.}~\bibnamefont{Ring}}, \bibinfo{journal}{Prog.
  Part. Nucl. Phys.} \textbf{\bibinfo{volume}{37}}, \bibinfo{pages}{193}
  (\bibinfo{year}{1996}).

\bibitem{Vretenar2005}
\bibinfo{author}{\bibfnamefont{D.}~\bibnamefont{Vretenar}},
  \bibinfo{author}{\bibfnamefont{A.}~\bibnamefont{Afanasjev}},
  \bibinfo{author}{\bibfnamefont{G.}~\bibnamefont{Lalazissis}},
  \bibnamefont{and} \bibinfo{author}{\bibfnamefont{P.}~\bibnamefont{Ring}},
  \bibinfo{journal}{Phys. Rep.} \textbf{\bibinfo{volume}{409}},
  \bibinfo{pages}{101} (\bibinfo{year}{2005}).

\bibitem{Meng2006}
\bibinfo{author}{\bibfnamefont{J.}~\bibnamefont{Meng}},
  \bibinfo{author}{\bibfnamefont{H.}~\bibnamefont{Toki}},
  \bibinfo{author}{\bibfnamefont{S.}~\bibnamefont{Zhou}},
  \bibinfo{author}{\bibfnamefont{S.}~\bibnamefont{Zhang}},
  \bibinfo{author}{\bibfnamefont{W.}~\bibnamefont{Long}}, \bibnamefont{and}
  \bibinfo{author}{\bibfnamefont{L.}~\bibnamefont{Geng}},
  \bibinfo{journal}{Prog. Part. Nucl. Phys.} \textbf{\bibinfo{volume}{57}},
  \bibinfo{pages}{470} (\bibinfo{year}{2006}).

\bibitem{Miller1975}
\bibinfo{author}{\bibfnamefont{L.~D.} \bibnamefont{Miller}},
  \bibinfo{journal}{Ann. Phys.} \textbf{\bibinfo{volume}{91}},
  \bibinfo{pages}{40} (\bibinfo{year}{1975}).

\bibitem{Serot1981}
\bibinfo{author}{\bibfnamefont{B.~D.} \bibnamefont{Serot}},
  \bibinfo{journal}{Phys. Lett. B} \textbf{\bibinfo{volume}{107}},
  \bibinfo{pages}{263} (\bibinfo{year}{1981}).

\bibitem{McNeil1986}
\bibinfo{author}{\bibfnamefont{J.~A.} \bibnamefont{McNeil}},
  \bibinfo{author}{\bibfnamefont{R.~D.} \bibnamefont{Amado}},
  \bibinfo{author}{\bibfnamefont{C.~J.} \bibnamefont{Horowitz}},
  \bibinfo{author}{\bibfnamefont{M.}~\bibnamefont{Oka}},
  \bibinfo{author}{\bibfnamefont{J.~R.} \bibnamefont{Shepard}},
  \bibnamefont{and} \bibinfo{author}{\bibfnamefont{D.~A.}
  \bibnamefont{Sparrow}}, \bibinfo{journal}{Phys. Rev. C}
  \textbf{\bibinfo{volume}{34}}, \bibinfo{pages}{746} (\bibinfo{year}{1986}).

\bibitem{Serot1992}
\bibinfo{author}{\bibfnamefont{B.~D.} \bibnamefont{Serot}},
  \bibinfo{journal}{Rep. Prog. Phys.} \textbf{\bibinfo{volume}{55}},
  \bibinfo{pages}{1855} (\bibinfo{year}{1992}).

\bibitem{Ichii1987a}
\bibinfo{author}{\bibfnamefont{S.}~\bibnamefont{Ichii}},
  \bibinfo{author}{\bibfnamefont{W.}~\bibnamefont{Bentz}},
  \bibinfo{author}{\bibfnamefont{A.}~\bibnamefont{Arima}}, \bibnamefont{and}
  \bibinfo{author}{\bibfnamefont{T.}~\bibnamefont{Suzuki}},
  \bibinfo{journal}{Phys. Lett. B} \textbf{\bibinfo{volume}{192}},
  \bibinfo{pages}{11} (\bibinfo{year}{1987}).

\bibitem{Shepard1988}
\bibinfo{author}{\bibfnamefont{J.~R.} \bibnamefont{Shepard}},
  \bibinfo{author}{\bibfnamefont{E.}~\bibnamefont{Rost}},
  \bibinfo{author}{\bibfnamefont{C.-Y.} \bibnamefont{Cheung}},
  \bibnamefont{and} \bibinfo{author}{\bibfnamefont{J.~A.}
  \bibnamefont{Mc~Neil}}, \bibinfo{journal}{Phys. Rev. C}
  \textbf{\bibinfo{volume}{37}}, \bibinfo{pages}{1130} (\bibinfo{year}{1988}).

\bibitem{Furnstahl1988}
\bibinfo{author}{\bibfnamefont{R.~J.} \bibnamefont{Furnstahl}},
  \bibinfo{journal}{Phys. Rev. C} \textbf{\bibinfo{volume}{38}},
  \bibinfo{pages}{370} (\bibinfo{year}{1988}).

\bibitem{Hofmann1988}
\bibinfo{author}{\bibfnamefont{U.}~\bibnamefont{Hofmann}} \bibnamefont{and}
  \bibinfo{author}{\bibfnamefont{P.}~\bibnamefont{Ring}},
  \bibinfo{journal}{Phys. Lett. B} \textbf{\bibinfo{volume}{214}},
  \bibinfo{pages}{307} (\bibinfo{year}{1988}).

\bibitem{Furnstahl1989}
\bibinfo{author}{\bibfnamefont{R.~J.} \bibnamefont{Furnstahl}}
  \bibnamefont{and} \bibinfo{author}{\bibfnamefont{C.~E.} \bibnamefont{Price}},
  \bibinfo{journal}{Phys. Rev. C} \textbf{\bibinfo{volume}{40}},
  \bibinfo{pages}{1398} (\bibinfo{year}{1989}).

\bibitem{Yao2006}
\bibinfo{author}{\bibfnamefont{J.~M.} \bibnamefont{Yao}},
  \bibinfo{author}{\bibfnamefont{H.}~\bibnamefont{Chen}}, \bibnamefont{and}
  \bibinfo{author}{\bibfnamefont{J.}~\bibnamefont{Meng}},
  \bibinfo{journal}{Phys. Rev. C} \textbf{\bibinfo{volume}{74}},
  \bibinfo{pages}{024307} (\bibinfo{year}{2006}).

\bibitem{Li2009}
\bibinfo{author}{\bibfnamefont{J.}~\bibnamefont{Li}},
  \bibinfo{author}{\bibfnamefont{Y.}~\bibnamefont{Zhang}},
  \bibinfo{author}{\bibfnamefont{J.~M.} \bibnamefont{Yao}}, \bibnamefont{and}
  \bibinfo{author}{\bibfnamefont{J.}~\bibnamefont{Meng}},
  \bibinfo{journal}{Sci. China Ser. G} \textbf{\bibinfo{volume}{52}},
  \bibinfo{pages}{1586} (\bibinfo{year}{2009}{\natexlab{a}}).

\bibitem{Li2009b}
\bibinfo{author}{\bibfnamefont{J.}~\bibnamefont{Li}},
  \bibinfo{author}{\bibfnamefont{J.~M.} \bibnamefont{Yao}}, \bibnamefont{and}
  \bibinfo{author}{\bibfnamefont{J.}~\bibnamefont{Meng}},
  \bibinfo{journal}{Chin. Phys. C} \textbf{\bibinfo{volume}{33 (S1)}},
  \bibinfo{pages}{98} (\bibinfo{year}{2009}{\natexlab{b}}).


\bibitem{Morse1990}
\bibinfo{author}{\bibfnamefont{T.~M.} \bibnamefont{Morse}},
  \bibinfo{author}{\bibfnamefont{C.~E.} \bibnamefont{Price}}, \bibnamefont{and}
  \bibinfo{author}{\bibfnamefont{J.~R.} \bibnamefont{Shepard}},
  \bibinfo{journal}{Phys. Lett. B} \textbf{\bibinfo{volume}{251}},
  \bibinfo{pages}{241} (\bibinfo{year}{1990}).

\bibitem{Li2011b}
\bibinfo{author}{\bibfnamefont{J.}~\bibnamefont{Li}},
  \bibinfo{author}{\bibfnamefont{J.~M.} \bibnamefont{Yao}},
  \bibinfo{author}{\bibfnamefont{J.}~\bibnamefont{Meng}}, \bibnamefont{and}
  \bibinfo{author}{\bibfnamefont{A.}~\bibnamefont{Arima}},
  \bibinfo{journal}{Prog. Theor. Phys.} \textbf{\bibinfo{volume}{125}},
  \bibinfo{pages}{1185} (\bibinfo{year}{2011}{\natexlab{a}}).

\bibitem{Li2011}
\bibinfo{author}{\bibfnamefont{J.}~\bibnamefont{Li}},
  \bibinfo{author}{\bibfnamefont{J.}~\bibnamefont{Meng}},
  \bibinfo{author}{\bibfnamefont{P.}~\bibnamefont{Ring}},
  \bibinfo{author}{\bibfnamefont{J.~M.} \bibnamefont{Yao}}, \bibnamefont{and}
  \bibinfo{author}{\bibfnamefont{A.}~\bibnamefont{Arima}},
  \bibinfo{journal}{Sci. China Phys. Mech. Astron.}
  \textbf{\bibinfo{volume}{54}}, \bibinfo{pages}{204}
  (\bibinfo{year}{2011}{\natexlab{b}}).

\bibitem{Wei2012}
\bibinfo{author}{\bibfnamefont{J.}~\bibnamefont{Wei}},
  \bibinfo{author}{\bibfnamefont{J.}~\bibnamefont{Li}}, \bibnamefont{and}
  \bibinfo{author}{\bibfnamefont{J.}~\bibnamefont{Meng}},
  \bibinfo{journal}{Prog. Theor. Phys. Supplement}
  \textbf{\bibinfo{volume}{196}}, \bibinfo{pages}{400} (\bibinfo{year}{2012}).

\bibitem{Furnstahl1987}
\bibinfo{author}{\bibfnamefont{R.~J.} \bibnamefont{Furnstahl}}
  \bibnamefont{and} \bibinfo{author}{\bibfnamefont{S.}~\bibnamefont{Brian~D.}},
  \bibinfo{journal}{Nucl. Phys. A} \textbf{\bibinfo{volume}{468}},
  \bibinfo{pages}{539} (\bibinfo{year}{1987}).

\bibitem{Buvenich2002}
\bibinfo{author}{\bibfnamefont{T.}~\bibnamefont{B\"{u}venich}},
  \bibinfo{author}{\bibfnamefont{D.~G.} \bibnamefont{Madland}},
  \bibinfo{author}{\bibfnamefont{J.~A.} \bibnamefont{Maruhn}},
  \bibnamefont{and} \bibinfo{author}{\bibfnamefont{P.-G.}
  \bibnamefont{Reinhard}}, \bibinfo{journal}{Phys. Rev. C}
  \textbf{\bibinfo{volume}{65}}, \bibinfo{pages}{044308}
  (\bibinfo{year}{2002}).

\bibitem{Niksic2005}
\bibinfo{author}{\bibfnamefont{T.}~\bibnamefont{Nik\ifmmode \check{s}\else
  \v{s}\fi{}i\ifmmode~\acute{c}\else \'{c}\fi{}}},
  \bibinfo{author}{\bibfnamefont{D.}~\bibnamefont{Vretenar}}, \bibnamefont{and}
  \bibinfo{author}{\bibfnamefont{P.}~\bibnamefont{Ring}},
  \bibinfo{journal}{Phys. Rev. C} \textbf{\bibinfo{volume}{72}},
  \bibinfo{pages}{014312} (\bibinfo{year}{2005}).

\bibitem{Daoutidis2009}
\bibinfo{author}{\bibfnamefont{J.}~\bibnamefont{Daoutidis}} \bibnamefont{and}
  \bibinfo{author}{\bibfnamefont{P.}~\bibnamefont{Ring}},
  \bibinfo{journal}{Phys. Rev. C} \textbf{\bibinfo{volume}{80}},
  \bibinfo{pages}{024309} (\bibinfo{year}{2009}).


\bibitem{DeConti1998}
\bibinfo{author}{\bibfnamefont{C.}~\bibnamefont{De Conti}},
  \bibinfo{author}{\bibfnamefont{A. P.}~\bibnamefont{Gale{\~a}o}}, \bibnamefont{and}
  \bibinfo{author}{\bibfnamefont{F.}~\bibnamefont{Krmpoti{\'c}}},
  \bibinfo{journal}{Phys. Lett. B} \textbf{\bibinfo{volume}{444}},
  \bibinfo{pages}{14} (\bibinfo{year}{1998}).

\bibitem{Paar2004}
\bibinfo{author}{\bibfnamefont{N.}~\bibnamefont{Paar}},
  \bibinfo{author}{\bibfnamefont{T.}~\bibnamefont{Nik\v{s}i\'{c}}},
  \bibinfo{author}{\bibfnamefont{D.}~\bibnamefont{Vretenar}}, \bibnamefont{and}
  \bibinfo{author}{\bibfnamefont{P.}~\bibnamefont{Ring}},
  \bibinfo{journal}{Phys. Rev. C} \textbf{\bibinfo{volume}{69}},
  \bibinfo{pages}{054303} (\bibinfo{year}{2004}).

\bibitem{Zhao2010}
\bibinfo{author}{\bibfnamefont{P.~W.} \bibnamefont{Zhao}},
  \bibinfo{author}{\bibfnamefont{Z.~P.} \bibnamefont{Li}},
  \bibinfo{author}{\bibfnamefont{J.~M.} \bibnamefont{Yao}}, \bibnamefont{and}
  \bibinfo{author}{\bibfnamefont{J.}~\bibnamefont{Meng}},
  \bibinfo{journal}{Phys. Rev. C} \textbf{\bibinfo{volume}{82}},
  \bibinfo{pages}{054319} (\bibinfo{year}{2010}).

\bibitem{Koepf1989}
\bibinfo{author}{\bibfnamefont{W.} \bibnamefont{Koepf}} \bibnamefont{and}
  \bibinfo{author}{\bibfnamefont{P.} \bibnamefont{Ring}},
  \bibinfo{journal}{Nucl. Phys. A} \textbf{\bibinfo{volume}{493}},
  \bibinfo{pages}{61} (\bibinfo{year}{1989}).


\bibitem{Green1965}
\bibinfo{author}{\bibfnamefont{A. M.} \bibnamefont{Green}},
  \bibinfo{author}{\bibfnamefont{A.} \bibnamefont{Kallio}}, \bibnamefont{and}
  \bibinfo{author}{\bibfnamefont{K.} \bibnamefont{Kolltveit}},
  \bibinfo{journal}{Phys. Lett.} \textbf{\bibinfo{volume}{14}},
  \bibinfo{pages}{142} (\bibinfo{year}{1965}).


\bibitem{Gillet1964}
\bibinfo{author}{\bibfnamefont{V.} \bibnamefont{Gillet}},
  \bibinfo{author}{\bibfnamefont{A. M.} \bibnamefont{Green}}, \bibnamefont{and}
  \bibinfo{author}{\bibfnamefont{E. A.} \bibnamefont{Sanderson}},
  \bibinfo{journal}{Phys. Lett.} \textbf{\bibinfo{volume}{11}},
  \bibinfo{pages}{44} (\bibinfo{year}{1964}).


\bibitem{Kim1963}
\bibinfo{author}{\bibfnamefont{Y. E.} \bibnamefont{Kim}} \bibnamefont{and}
  \bibinfo{author}{\bibfnamefont{J. O.} \bibnamefont{Rasmussen}},
  \bibinfo{journal}{Nucl. Phys.} \textbf{\bibinfo{volume}{47}},
  \bibinfo{pages}{184} (\bibinfo{year}{1963}).


\bibitem{Brueckner1958}
\bibinfo{author}{\bibfnamefont{K. A.} \bibnamefont{Brueckner}} \bibnamefont{and}
  \bibinfo{author}{\bibfnamefont{J. L.} \bibnamefont{Gammel}},
  \bibinfo{journal}{Phys. Rev.} \textbf{\bibinfo{volume}{109}},
  \bibinfo{pages}{1023} (\bibinfo{year}{1958}).


\bibitem{Hamada1962}
\bibinfo{author}{\bibfnamefont{T.} \bibnamefont{Hamada}} \bibnamefont{and}
  \bibinfo{author}{\bibfnamefont{I. D.} \bibnamefont{Johnston}},
  \bibinfo{journal}{Nucl. Phys.} \textbf{\bibinfo{volume}{34}},
  \bibinfo{pages}{382} (\bibinfo{year}{1962}).

\bibitem{Bertsch1977}
\bibinfo{author}{\bibfnamefont{G.} \bibnamefont{Bertsch}},
  \bibinfo{author}{\bibfnamefont{J.} \bibnamefont{ Borysowicz}},
  \bibinfo{author}{\bibfnamefont{H.} \bibnamefont{ McManus}} \bibnamefont{and}
  \bibinfo{author}{\bibfnamefont{W. G.} \bibnamefont{Love}},
  \bibinfo{journal}{Nucl. Phys. A} \textbf{\bibinfo{volume}{284}},
  \bibinfo{pages}{399} (\bibinfo{year}{1977}).


\bibitem{Arima1972}
\bibinfo{author}{\bibfnamefont{A.}~\bibnamefont{Arima}} \bibnamefont{and}
  \bibinfo{author}{\bibfnamefont{L.~J.} \bibnamefont{Huang-Lin}},
  \bibinfo{journal}{Phys. Lett. B} \textbf{\bibinfo{volume}{41}},
  \bibinfo{pages}{435} (\bibinfo{year}{1972}).


\bibitem{Blomqvist1965}
\bibinfo{author}{\bibfnamefont{J.} \bibnamefont{Blomqvist}},
  \bibinfo{author}{\bibfnamefont{N.}~\bibnamefont{Freed}}, \bibnamefont{and}
  \bibinfo{author}{\bibfnamefont{H. O.} \bibnamefont{Zetterstr{\"o}m}},
  \bibinfo{journal}{Phys. Lett.} \textbf{\bibinfo{volume}{18}},
  \bibinfo{pages}{47} (\bibinfo{year}{1965}).




\bibitem{Ring1973}
\bibinfo{author}{\bibfnamefont{P.}~\bibnamefont{Ring}} \bibnamefont{and}
  \bibinfo{author}{\bibfnamefont{E.} \bibnamefont{Werner}},
  \bibinfo{journal}{Nucl. Phys. A} \textbf{\bibinfo{volume}{211}},
  \bibinfo{pages}{198} (\bibinfo{year}{1973}).


\bibitem{Bernard1980}
\bibinfo{author}{\bibfnamefont{V.}~\bibnamefont{Bernard}} \bibnamefont{and}
  \bibinfo{author}{\bibfnamefont{N.} \bibnamefont{Van Giai}},
  \bibinfo{journal}{Nucl. Phys. A} \textbf{\bibinfo{volume}{348}},
  \bibinfo{pages}{75} (\bibinfo{year}{1980}).


\bibitem{Litvinova2006}
\bibinfo{author}{\bibfnamefont{E.}~\bibnamefont{Litvinova}} \bibnamefont{and}
  \bibinfo{author}{\bibfnamefont{P.} \bibnamefont{Ring}},
  \bibinfo{journal}{Phys. Rev. C} \textbf{\bibinfo{volume}{73}},
  \bibinfo{pages}{044328} (\bibinfo{year}{2006}).

\bibitem{Ring2009}
\bibinfo{author}{\bibfnamefont{P.} \bibnamefont{Ring}} \bibnamefont{and}
\bibinfo{author}{\bibfnamefont{E.}~\bibnamefont{Litvinova}},
  \bibinfo{journal}{Phys. At. Nucl.} \textbf{\bibinfo{volume}{72}},
  \bibinfo{pages}{1285} (\bibinfo{year}{2009})



\bibitem{Litvinova2007}
\bibinfo{author}{\bibfnamefont{E.} \bibnamefont{Litvinova}},
\bibinfo{author}{\bibfnamefont{P.}~\bibnamefont{Ring}}, \bibnamefont{and}
\bibinfo{author}{\bibfnamefont{V.} \bibnamefont{Tselyaev}}
  \bibinfo{journal}{Phys. Rev. C} \textbf{\bibinfo{volume}{75}},
  \bibinfo{pages}{064308} (\bibinfo{year}{2007})


\bibitem{Moghrabi2012}
\bibinfo{author}{\bibfnamefont{K.} \bibnamefont{Moghrabi}} \bibnamefont{and}
\bibinfo{author}{\bibfnamefont{M.}~\bibnamefont{Grasso}},
  \bibinfo{journal}{Phys. Rev. C} \textbf{\bibinfo{volume}{86}},
  \bibinfo{pages}{044319} (\bibinfo{year}{2012})

\bibitem{Moghrabi2012a}
\bibinfo{author}{\bibfnamefont{K.} \bibnamefont{Moghrabi}},
\bibinfo{author}{\bibfnamefont{M.}~\bibnamefont{Grasso}},
\bibinfo{author}{\bibfnamefont{X.} \bibnamefont{Roca-Maza}}, \bibnamefont{and}
\bibinfo{author}{\bibfnamefont{G.} \bibnamefont{Col\`o}},
  \bibinfo{journal}{Phys. Rev. C} \textbf{\bibinfo{volume}{85}},
  \bibinfo{pages}{044323} (\bibinfo{year}{2012})


\bibitem{Bauer1973}
\bibinfo{author}{\bibfnamefont{R.}~\bibnamefont{Bauer}},
  \bibinfo{author}{\bibfnamefont{J.}~\bibnamefont{Speth}},
  \bibinfo{author}{\bibfnamefont{V.}~\bibnamefont{Klemt}},
  \bibinfo{author}{\bibfnamefont{P.}~\bibnamefont{Ring}},
  \bibinfo{author}{\bibfnamefont{E.}~\bibnamefont{Werner}}, \bibnamefont{and}
  \bibinfo{author}{\bibfnamefont{T.}~\bibnamefont{Yamazaki}},
  \bibinfo{journal}{Nucl. Phys. A} \textbf{\bibinfo{volume}{209}},
  \bibinfo{pages}{535} (\bibinfo{year}{1973}).

\bibitem{Rho1974}
\bibinfo{author}{\bibfnamefont{Mannque}~\bibnamefont{Rho}},
  \bibinfo{journal}{Nucl. Phys. A} \textbf{\bibinfo{volume}{231}},
  \bibinfo{pages}{493} (\bibinfo{year}{1974}).

\bibitem{Oset1979}
\bibinfo{author}{\bibfnamefont{E.} \bibnamefont{Oset}}  \bibnamefont{and}
  \bibinfo{author}{\bibfnamefont{M.}~\bibnamefont{Rho}},
  \bibinfo{journal}{Phys. Rev. Lett.} \textbf{\bibinfo{volume}{42}},
  \bibinfo{pages}{47} (\bibinfo{year}{1979}).


\bibitem{Knupfer1980}
\bibinfo{author}{\bibfnamefont{W.} \bibnamefont{Kn{\"u}pfer}},
  \bibinfo{author}{\bibfnamefont{M.}~\bibnamefont{Dillig}}, \bibnamefont{and}
  \bibinfo{author}{\bibfnamefont{A.}~\bibnamefont{Richter}},
  \bibinfo{journal}{Phys. Lett. B} \textbf{\bibinfo{volume}{95}},
  \bibinfo{pages}{349} (\bibinfo{year}{1980}).


\bibitem{Long2006}
\bibinfo{author}{\bibfnamefont{W.-H.} \bibnamefont{Long}},
  \bibinfo{author}{\bibfnamefont{N.}~\bibnamefont{Van~Giai}}, \bibnamefont{and}
  \bibinfo{author}{\bibfnamefont{J.}~\bibnamefont{Meng}},
  \bibinfo{journal}{Phys. Lett. B} \textbf{\bibinfo{volume}{640}},
  \bibinfo{pages}{150} (\bibinfo{year}{2006}).

\bibitem{Mattuck1992}
\bibinfo{author}{\bibfnamefont{R.~D.} \bibnamefont{Mattuck}},
  \emph{\bibinfo{title}{A Guide to Feynman Diagrams in the Many-Body Problem}}
  (\bibinfo{publisher}{Dover Publications}, \bibinfo{year}{1992}).




\end{thebibliography}

\end{document}